\documentclass[letterpaper, 10pt, twocolumn, pra]{revtex4}
\usepackage{amsmath}
\usepackage{graphicx}
\usepackage{empheq}
\begin{document}
\title{A Canonical Transformation Theory from Extended Normal Ordering}
\author{Takeshi Yanai\footnote{Present address: Department of Theoretical and Computational Molecular Science, Institute for Molecular Science, Okazaki, Aichi 444-8585, Japan. Electronic mail: yanait@ims.ac.jp} and Garnet Kin-Lic Chan}
\affiliation{Department of Chemistry and Chemical Biology\\Cornell
  University, Ithaca, NY 14853-1301}
\date{\today}

\newcommand{\Ehartree}{{$E_\mathrm{h}$}}
\newcommand{\MK}{{MK}}

\renewcommand{\topfraction}{1.0}
\renewcommand{\bottomfraction}{1.0}
\renewcommand{\textfraction}{0.0}

\begin{abstract}
The Canonical Transformation theory of Yanai and Chan [J.\ Chem.\ Phys.\ {\bf 124}, 194106 (2006)]
provides a rigorously size-extensive description of dynamical
correlation in multireference problems.  Here we describe a
new formulation of the theory based on the extended normal
ordering procedure of Mukherjee and Kutzelnigg [J.\ Chem.\ Phys.\ {\bf 107}, 432 (1997)].
On studies of the water, nitrogen, and iron-oxide potential energy
curves, the Linearised Canonical Transformation Singles and Doubles
theory is competitive in accuracy with some of the best multireference
methods, such as the Multireference Averaged Coupled Pair Functional,
while computational timings (in the case of the iron-oxide molecule)
are two-three orders of magnitude faster and comparable to those of
Complete Active Space Second-Order Perturbation Theory. The results
presented here are greatly improved both in accuracy and in cost over
our earlier study as the result of a new numerical algorithm for
solving the amplitude equations.
\end{abstract}
\maketitle

\section{Introduction}

While nondynamic correlation between electrons establishes the
qualitative features of chemical bonding, it is the accurate
description of dynamic correlation, associated with the short-range
cusp behaviour of the wavefunction, which is necessary to obtain
quantitative agreement with experiment. Starting from a suitable
reference function, the exponential ansatz provides an accurate and
economical description of dynamic correlation. For example, in systems
that are qualitatively described by a single determinant reference,
Coupled Cluster (CC) theory paired with a large basis set yields
predictions with chemical accuracy \cite{BAR-STA:1994:_cc_review, HRJOK:2004:_cc_review, LEE-SCU:1995:_cc_review}.
However, for the many chemical problems which require a multireference
characterisation, a practical theory for dynamic correlation with the
desirable qualities of the exponential ansatz - size-extensivity,
chemical-accuracy, and moderate computational cost - has yet to be
widely established.

In an earlier article \cite{YAN-CHA:2006:_ctpaper} we presented a Canonical
Transformation (CT) theory which is based on an exponential ansatz, is
rigorously size-extensive, and which may easily be combined with any
multireference starting wavefunction.
In the form implemented in that work the computational cost is
$O(a^2 e^4)$, where $a$ is the number of active orbitals and $e$ is the
number of external orbitals.  In calculations of bond-breaking
potential energy curves, the Linearised Canonical Transformation
Doubles (L-CTD) theory performed significantly better than
multireference perturbation theory, and obtained the accuracy of
Coupled Cluster Single Doubles (CCSD) at the equilibrium geometry
across the \textit{entire} potential energy curve. Our work was
directly motivated by the Canonical Diagonalisation theory of White
\cite{WHITE:2002:_ct} although there are earlier related contributions as we
describe below.

The purpose of the current work is to improve on our initial
contribution in several areas. A central feature of the Canonical
Transformation theory is the use of an \textit{operator decomposition},
both to close the infinite expansions associated
with an exponential ansatz and to reduce the complexity of the energy
and amplitude equations that arise when working with a complicated
reference function.  In our earlier work, we introduced a
cumulant-type operator decomposition by analogy to the cumulant
decomposition of density matrices found in reduced density matrix
theories \cite{COL-VAL:1993:_cse, COL-VAL:1994:_cse, NAK-YAS:1996:_cse, YAS-NAK:1997:_cse, MAZ:1998:_cse1, MAZ:1998:_cse2}.
However, this choice of operator
decomposition is not unique and here we explore an alternative
operator decomposition, with some formal advantages, that is based on
the concept of \textit{extended normal ordering} as introduced by
Mukherjee and Kutzelnigg \cite{MUKHERJEE:1995:_normalorder, MUKHERJEE:1997:_normalorder, KUT-MUK:1997:_normalorder}.
Indeed, examination of the articles by these authors shows that they
anticipated the utility of their results in multireference correlation
theories, and in this context our current theory is in part a
realisation along such directions.

A second focus of this work is to investigate in detail the behaviour
of the Canonical Transformation theory in a variety of chemical
problems. For example we study, with a range of basis sets, the
bond-breaking potential energy curves of water, nitrogen, and
iron-oxide and compare our results against state-of-the-art
multi-reference configuration interaction and perturbation
theories. In addition, we examine numerically the size-extensivity and
density-scaling properties of the Canonical Transformation
energies. The results in the present study are much improved over our
earlier work, in large part because of improvements we have made to
our numerical algorithms, and we describe in detail the numerical
aspects of efficiently implementing and converging the CT equations.




\section{Canonical Transformation Theory}

\label{sec:ct-theory}
\subsection{Recapitulation}

In multireference problems we divide the orbitals into
\textit{active orbitals}, which describe the nondynamic correlation and
\textit{external orbitals} which describe the dynamic correlation. The external
orbitals may further be divided into core and virtual orbitals; core
orbitals are those which remain doubly occupied in all the reference
configurations.

We will assume that a reference wavefunction $\Psi_0$
is available that describes the nondynamic correlation in the
problem. This may be obtained, for example, from a
Complete Active Space Self-Consistent Field (CASSCF) calculation that
exactly correlates electrons within the active orbitals
\cite{ROOS:1987:_casscf, RSGE:1982:_casscf}.
Alternatively, and especially for larger active spaces, a Density Matrix Renormalization Group
wavefunction may be used \cite{CHAN-HEA:2002:_dmrg, Hachmann2006}.
We then incorporate the remaining dynamic correlation on top of the reference wavefunction
$\Psi_0$ via an exponential operator that generates excitations
between the active and external spaces yielding
\begin{equation}
\Psi = e^A \Psi_0 \label{eq:expansatz}
\end{equation}
We will be concerned with a \textit{unitary} formulation, where $A^\dag =
-A$. The excitations are understood to be both of external and
semi-internal form
\begin{equation}
A = A^a_i (a^a_i - a^i_a)
+ A^{ab}_{ij} (a^{ab}_{ij} - a^{ij}_{ab})
+ A^{ak}_{ij} (a^{ak}_{ij} - a^{ij}_{ak})
+ \ldots \label{eq:aoperator}
\end{equation}
where $ijk \ldots$ denote active indices, $abc \ldots$, external
indices, $a^a_i = a_a^\dag a_i,\,\, a^{ab}_{ij} = a^\dag_a a^\dag_b a_j
a_i$, and the summation convention is assumed. For example, the first
two terms are the usual external single and double excitations, while the third
term (with three active indices) is a
semi-internal single excitation, which captures the coupling between
singles relaxation in the active space and singles excitation to the
external space.

In a related picture, we can also view $e^A$ as generating an effective
\textit{canonically transformed} Hamiltonian $\bar{H}$ that acts only in the active space, but which
has dynamic correlation folded in from the external space, where
\begin{align}
\bar{H} &= e^{-A} H  e^A \\
\bar{H} \Psi &= E \Psi \label{eq:schrodeq}
\end{align}

The exponential ansatz combined with a multireference wavefunction
$\Psi_0$ as shown in eqn.\ (\ref{eq:expansatz}) has a long history and
we necessarily can only give an incomplete account here.  Such an
ansatz is used in some forms of multi-reference coupled cluster theory
(MRCC) as discussed in the review by Paldus and Li
\cite{PAL-LI:1999:_cc_mrcc}. In particular, an early example of a
complete theoretical scheme for a related multi-reference coupled
cluster method  was  given by Mukherjee  
in Ref. \cite{MUKHERJEE:1995:_normalorder}.
While CC theory is usually formulated in terms of
similarity rather than canonical (i.e.\ unitary) transforms, unitary
exponentials have previously been explored in a multi-reference
setting by Freed et al \cite{FREED:1989:_mrpt}, Kirtman et al
\cite{KIRTMAN:1982:_qdpt}, and Simons et al
\cite{HOF-SIM:1988:_mrucc}. We mention also the single-reference
unitary coupled cluster work by Kutzelnigg
\cite{KUTZ:1982:_ucc,KUTZ:1984:_ucc}, Bartlett et al
\cite{WTB:1989:_ucc, BWN:1989:_ucc,TAU-BAR:2006:_ucc}, and Pal
\cite{PAL:1983:_ucc,PAL:1984:_ucc}.  The general concept of effective
Hamiltonians and canonical transformations is of course very old,
dating back to van Vleck \cite{VLECK:1929:_qdpt}. We note in particular some
modern theories that emphasize an effective Hamiltonian language
similar to our own such as the Effective Valence Hamiltonian theory of
Freed \cite{FREED:1989:_mrpt} and the Generalized van Vleck theory of
Kirtman \cite{KIRTMAN:1982:_qdpt}. As recognised by Freed, the folding
in of dynamic correlation into the active-space effective Hamiltonian
is a form of renormalisation transformation. This picture was pursued
by White in his theory of Canonical Diagonalisation
\cite{WHITE:2002:_ct}, and as described previously, this is the
primary precursor to our work.

In the exponential ansatz of single-reference coupled cluster theory,
the commutativity of the excitation operators in the
single-reference form of $A$, i.e.\ $A=A^a_i a^a_i + A^{ab}_{ij}
a^{ab}_{ij}$, allows the Baker-Campbell-Hausdorff expansion of $\bar{H}$ to terminate at
low-order for low-particle rank in $A$.
The difficulty in working with the  multireference
exponential ansatz arises from the non-commuting excitations in the
multi-reference form of $A$ in eqn.\ (\ref{eq:aoperator}), which leads to a non-terminating expansion for the  effective
Hamiltonian $\bar{H}$. (In fact this difficulty already arises if we
use  a \textit{unitary} $e^A$ with the single-reference form of $A$).


In our earlier Canonical Transformation (CT) theory we introduced a
new route to a tractable and computationally efficient formulation for
the multireference ansatz (\ref{eq:expansatz}). 
Starting
from the Baker-Campbell-Hausdorff expansion of the exact effective
Hamiltonian,
\begin{equation}
\bar{H} = H + [H, A] + \frac{1}{2} [[H, A], A] + \ldots \label{eq:origbch}
\end{equation}
we replace each commutator by an approximate \textit{decomposed} commutator, to
yield an approximate effective Hamiltonian
\begin{equation}
\bar{H}_{1, 2, \ldots} = H + [H, A]_{1,2, \ldots} + \frac{1}{2} [[H,
A]_{1, 2, \ldots}, A]]_{1, 2, \ldots} + \ldots \label{eq:decompbch}
\end{equation}
Each subscript denotes a decomposition, and the numbers $1, 2 \ldots$ denote
the particle ranks of the operators that remain after the
decomposition. 
Note that if \textit{all} particle ranks were included in
the decomposition (i.e. the
subscripts ranged from $1, 2, \ldots n$, where $n$ is the number of particles), then eqns. (\ref{eq:origbch}) and
(\ref{eq:decompbch}) would be identical. If in addition to including
all particle ranks in eqn. (\ref{eq:decompbch}) $A$  contained up to
$n$-body excitations, then
the CT ansatz (\ref{eq:expansatz}) would be exact in the sense of
full configuration interaction, and indeed eqn. (\ref{eq:schrodeq})
would hold exactly.  The two relevant approximations 
thus arise from restricting the excitations in $A$ (wavefunction ansatz) as well as the
form of the operator decomposition (operator ansatz) 
\footnote{Coupled cluster theory may similarly be viewed as involving two sets of approximations: truncation of the excitation operators, and approximate solution of the Schr\"odinger equation in a restricted projected space of excitations}.

As an example, let us consider  the linearised CT single and doubles
theory (L-CTSD) introduced in our earlier work. Here $A$ is restricted to contain only one- and
two-particle excitations  as in eqn. (\ref{eq:aoperator}), and we  restrict all decomposed commutators
to contain at most one and two-body operators (i.e.\ subscripts $1,
2$). Since $[H, A]$ generates a three-body operator, this requires some
decomposition of a three-body operator into lower body
operators. We proposed  an explicit decomposition into one-
and two-body operators based on
an analogy to the cumulant decomposition of density matrices,
\begin{align}
a^{pqr}_{stu}
\Rightarrow 9 (\gamma^p_s  \wedge a^{qr}_{tu})
          - 12 (  \gamma^p_s \wedge \gamma^q_t \wedge a^r_u)  \label{eq:decomp}
\end{align}
where in the above $\wedge$ denotes an antisymmetrisation over all upper
and lower indices with an associated factor of $1/(p\!)^2$, i.e.\ $1/36$ in the
above case, where $p$ is the particle rank of the original operator.
Here we will present the explicit steps leading to the above
decomposition.
Our notation follows closely that of Kutzelnigg and Mukherjee \cite{KUT-MUK:1999:_cumulant}.
Recall that the cumulant decomposition provides a way to
rewrite reduced density matrices $\gamma$ in terms of products of cumulants
$\lambda$, via
\begin{align}
\gamma^p_s &= \langle a^p_s \rangle  = \lambda^p_s \label{eq:onecum} \\
\gamma^{pq}_{st} &= \langle a^{pq}_{st} \rangle
 = \lambda^{pq}_{st} +
\gamma^p_s \gamma^q_t - \gamma^p_t \gamma^q_s \label{eq:twocum} \\
\gamma^{pqr}_{stu} &= \langle a^{pqr}_{stu} \rangle=
\lambda^{pqr}_{stu}
+ \gamma^p_s \lambda^{qr}_{tu} - \gamma^p_t \lambda^{qr}_{su} +
\gamma^p_u \lambda^{qr}_{st} \nonumber \\
- &\gamma^q_s \lambda^{pr}_{tu} + \gamma^q_t \lambda^{pr}_{su} -
\gamma^q_u \lambda^{pr}_{st} + \gamma^r_s \lambda^{pr}_{tu} - \gamma^r_t \lambda^{pr}_{su} +
\gamma^r_u \lambda^{pr}_{st} \nonumber \\
+ &\gamma^p_s \gamma^q_t \gamma^r_u
- \gamma^p_s \gamma^r_t \gamma^q_u
+ \gamma^q_s \gamma^r_t \gamma^p_u - \gamma^q_s \gamma^p_t
\gamma^r_u \nonumber \\
+ & \gamma^r_s \gamma^p_t \gamma^q_u
- \gamma^r_s \gamma^q_t \gamma^p_u  \label{eq:threecum}
\end{align}
For the three-particle density matrix, by dropping the three-particle
cumulant $\lambda^{pqr}_{stu}$, and substituting the
expressions (\ref{eq:onecum}) and (\ref{eq:twocum}) in (\ref{eq:threecum}), we obtain an
approximate decomposition in terms of one- and two-particle density
matrices only
\begin{align}
\gamma^{pqr}_{stu} & \Rightarrow
\gamma^p_s \gamma^{qr}_{tu} - \gamma^p_t \gamma^{qr}_{su} +
\gamma^p_u \gamma^{qr}_{st} \nonumber \\
- &\gamma^q_s \gamma^{pr}_{tu} + \gamma^q_t \gamma^{pr}_{su} -
\gamma^q_u \gamma^{pr}_{st} + \gamma^r_s \gamma^{pr}_{tu} - \gamma^r_t \gamma^{pr}_{su} +
\gamma^r_u \gamma^{pr}_{st} \nonumber \\
- &2(\gamma^p_s \gamma^q_t \gamma^r_u
- \gamma^p_s \gamma^r_t \gamma^q_u
+ \gamma^q_s \gamma^r_t \gamma^p_u
- \gamma^q_s \gamma^p_t \gamma^r_u\nonumber \\
+ &\gamma^r_s \gamma^p_t \gamma^q_u
- \gamma^r_s \gamma^q_t \gamma^p_u) \nonumber \\
&=9 (\gamma^p_s \wedge \gamma^{qr}_{tu}) - 12 (\gamma^p_s \wedge
\gamma^q_t \wedge \gamma^r_u) \label{eq:dendecomp}
\end{align}
To obtain our operator decomposition, we
simply replaced  expectation values in the above terms
by the corresponding operators, i.e.\ $\gamma^{pq}_{st} \to
a^{pq}_{st}$ and $\gamma^p_s \to a^p_s$, yielding eqn.\ (\ref{eq:decomp}).
Note that by construction, the expectation value of the operator decomposition reproduces the
three-particle density matrix cumulant decomposition (\ref{eq:dendecomp}).

By using this decomposition recursively i.e.\ by constructing the
double commutator by first using the decomposed  single commutator
$[H, A]_{1,2}$ as in eqn.\ (\ref{eq:decompbch}), the full effective
Hamiltonian $\bar{H}_{1, 2}$ at the L-CTSD level
\textit{contains only one and two-body operators}. Evaluation of the
energy then only requires the  one- and two-particle density matrices of
the reference function. As discussed in our initial work, this
fulfils one of the criteria for an efficient multireference theory,
namely, we do not need to explicitly manipulate the
complicated reference function. 
From a different perspective,
the canonical tranformations
can also be viewed as providing a parametrisation of a two-particle density
matrix theory. Recently, such connections have been explored from a
different direction by Mazziotti \cite{MAZZIOTTI:2006:_antiherm, MAZZIOTTI:2007:_antiherm}
and while interesting, we shall not dwell further on these matters here.

We call the above formulation a linearised theory, because the
operator decomposition is applied at the first commutator.
Then, at the L-CTSD level the energies and amplitudes are  evaluated
via
\begin{align}
E &= \langle \Psi_0 |\bar{H}_{1,2} |\Psi_0 \rangle  \\
0 &= \langle \Psi_0 |[\bar{H}_{1,2}, a^a_i - a^i_a]_{1,2} |\Psi_0 \rangle \\
0 &= \langle \Psi_0 |[\bar{H}_{1,2}, a^{ab}_{ij} - a^{ij}_{ab}]_{1,2} |\Psi_0 \rangle \\
0 &= \langle \Psi_0 |[\bar{H}_{1,2}, a^{ak}_{ij} - a^{ij}_{ak}]_{1,2} |\Psi_0 \rangle
\end{align}
The resulting computational cost of the theory is $O(a^2 e^4)$,
and is thus comparable to that of a \textit{single-reference} coupled cluster calculation.

\subsection{Accuracy of the operator decomposition}
\label{sec:accuracy_decomposition}

As presented above, the accuracy of the
Canonical Transformation theory rests on the accuracy of operator
decomposition,  given at the L-CTSD level by
eqn.\ (\ref{eq:decomp}). However, although  our  operator decomposition
was chosen so that its expectation value would  reproduce the density matrix cumulant
decomposition, \textit{this  choice is not unique}. For example, we could add to the r.h.s. of eqn.\ (\ref{eq:decomp}) any term with vanishing
expectation value with $\Psi_0$ and still preserve the correspondence
with the density matrix cumulant decomposition
(\ref{eq:dendecomp}). This simply reflects the fact that a
decomposition for expectation values (i.e. the cumulant decomposition)
does not contain sufficient information to specify a corresponding operator decomposition.

In our earlier work, we examined the accuracy of the operator
decomposition through a perturbative analysis of
 CT theory starting from a single determinantal
wavefunction $\Psi_D$ and using a single-reference single-doubles
excitation operator $A = A^a_i (a^a_i - a^i_a) + A^{ab}_{ij}
(a^{ab}_{ij} - a^{ij}_{ab})$. This analysis showed that the L-CTSD
theory was accurate through third-order in the fluctuation potential
$W = H - F$ where $F$ is the Fock operator i.e.\
\begin{align}
&\langle \Psi_D | \bar{H} | \Psi_D \rangle \nonumber\\
 =& \langle \Psi_D | \bar{H}_{1,2} |\Psi_D \rangle + O(W^4) \nonumber\\
 =& \langle \Psi_D |H + [H, A]_{1,2}+ [[H, A]_{1,2}, A]_{1,2} |\Psi_D \rangle + O(W^4)
\end{align}
However, consider what happens if we use the more general \textit{multireference}
form of $A$ in eqn.\ (\ref{eq:aoperator}) that includes semi-internal
excitations such as $A^{ak}_{ij} (a^{ak}_{ij} - a^{ij}_{ak})$,
together with a single reference wavefunction $|\Psi_D\rangle$.
Such excitations should not contribute as they destroy the single
reference wavefunction, and thus all expectation values of exact commutators
containing only semi-internal  excitations, e.g.
$\langle \Psi_D | [H, A^{ak}_{ij} (a^{ak}_{ij} - a^{ij}_{ak})] | \Psi_D\rangle$,
$\langle \Psi_D |[[H, A^{ak}_{ij} (a^{ak}_{ij} - a^{ij}_{ak})], A^{bn}_{lm} (a^{bn}_{lm} - a^{lm}_{bn})]|\Psi_D\rangle$
must vanish. However, using the cumulant-based operator
decomposition (\ref{eq:decomp})
we find that although the expectation value of the first commutator
$\langle \Psi_D|[H, A^{ak}_{ij} (a^{ak}_{ij} - a^{ij}_{ak})]_{1,2}|\Psi_D\rangle$,
correctly vanishes,
it does not do so for the second commutator. Non-vanishing terms arise e.g.\ from
\begin{equation}
\langle \Psi_D|[[H, A^{ak}_{ij} (a^{ak}_{ij} - a^{ij}_{ak})]_{1,2}, A^{bn}_{lm} (a^{bn}_{lm} - a^{lm}_{bn})] |\Psi_D\rangle\label{eq:nonvanishing_term}
\end{equation}
Writing $H$ and the two $A$ operators as $g^\dag g^\dag g g$, $o^\dag o^\dag ov$,
$v^\dag o^\dag o o$ respectively, using $g, o,
v$ to denote general, occupied, and virtual indices respectively, we
can see a non-zero contribution arising from
\begin{equation}
\langle \Psi_D | (g^\dag g^\dag g \underbracket{g) \ \ (o^\dag}  \overbracket{o^\dag o}
\underbracket{v) \ \ (v^\dag} \overbracket{o^\dag o}o) | \Psi_D \rangle \neq 0
\end{equation}
where the underbracket denotes contraction and the
overbracket denotes a replacement by a density matrix in the operator
decomposition.

In a multireference situation, we use the \textit{same} extended
excitation operator $A$ (with semi-internal excitations) through the entire potential energy surface,
even when the underlying reference wavefunction is largely of a single
reference nature, as is sometimes the case near the equilibrium
geometry. Thus the above deficiency of the cumulant operator
decomposition for single reference wavefunctions
motivates us to examine other possible
 decompositions, as we describe now.

\section{Extended Normal Ordering}

\subsection{Normal ordering for a Multireference wavefunction}

Normal ordering provides a standard way to decompose  an operator into
a sum of  zero-, one-, two- and higher body  contributions that are
ordered with respect to a given vacuum.
In many-body theory it is common to use normal ordering not with
respect to the physical vacuum, but rather with respect to a single
determinant state or Fermi
vacuum. With respect to the Fermi
vacuum, normal ordering of the operators $a^p_s, a^{pq}_{rs},
a^{pqr}_{stu}$ yields
\begin{align}
\tilde{a}^p_s & = a^p_s-\delta^p_s n_s \label{eq:fermione}\\
\tilde{a}^{pq}_{st} & = a^{pq}_{st} - \delta^p_s n_s a^q_t -
\delta^q_t n_t a^p_s \nonumber \\
& + \delta^p_t n_t a^q_s + \delta^q_s n_s a^p_t +
\delta^{pq}_{st} n_p n_q \label{eq:fermitwo} \\
\tilde{a}^{pqr}_{stu} & = a^{pqr}_{stu} -\delta^p_s n_p a^{qr}_{tu} +
\delta^q_s  n_q a^{pt}_{ru} + \delta^r_s n_r a^{qp}_{tu} \nonumber \\
&+  \delta^p_t n_p a^{qr}_{su} - \delta^q_t n_q a^{pr}_{su} + \delta^r_t n_r
a^{pq}_{su} + \delta^p_u n_p a^{qr}_{ts} \nonumber \\
&+  \delta q_u n_q a^{pr}_{st}
- \delta^r_u n_r a^{pq}_{st} + n_p n_q \delta^{pq}_{st} a^r_u + n_p
n_r \delta^{pr}_{su} a^q_t \nonumber \\
&+ n_q n_r \delta^{qr}_{tu} a^p_s - n_p n_q
\delta^{pq}_{su} a^r_t - n_p n_q \delta^{pq}_{ut} a^r_s - n_p n_r
\delta^{pr}_{st} a^q_u  \nonumber \\
&-n_p n_r \delta^{pr}_{tu} a^q_s  - n_q n_r
\delta^{qr}_{ts} a^p_u - n_q n_r \delta^{qr}_{su} a^p_t - n_p n_q n_r \delta^{pqr}_{stu}  \label{eq:threebodyfermidecomp}
\end{align}
where the tilde represents operators normal-ordered w.r.t. the Fermi
vacuum (quasi-particle operators), $n_p$ is the occupation number (0 or 1) of the $p$-th orbital, and $\delta^{pq}_{rs} = \delta^p_r \delta^q_s - \delta^p_s
\delta^q_r$, $\delta^{pqr}_{stu} = \delta^p_s \delta^q_r \delta^t_u +
\delta^p_t \delta^q_u \delta^r_s + \delta^p_u \delta^q_s \delta^r_t -
\delta^p_t \delta^q_s \delta^r_u - \delta^p_u \delta q_t \delta^r_s -
\delta^p_s \delta^q_u \delta^r_t$. Note that all normal-ordered
operators (other than the ``zero-body'' constant
term) yield  a vanishing expectation value with the Fermi vacuum,
e.g. $\langle \tilde{a}^p_s \rangle = 0$.
If we are interested
in  a state which is well approximated by
the Fermi vacuum, the higher-particle rank quasi-particle
operators such as $\tilde{a}^{pqr}_{stu}$ are less relevant to its properties than the lower-rank ones, since they
represent multiple simultaneous excitations away from the state. Thus the
Fermi-vacuum normal ordering presents  a natural way to approximate
high-particle rank operators   in terms of simpler lower-body terms by
simply neglecting the high particle-rank quasi-particle operators that appear in the normal
ordered form. For example, to approximate $a^{pqr}_{stu}$ in terms of
one- and two-body operators alone, we would  neglect
$\tilde{a}^{pqr}_{stu}$ in eqn.\ (\ref{eq:threebodyfermidecomp}).

In the Canonical Transformation theory, however, we are often
interested in reference states which cannot be represented well by any
Fermi vacuum. Recently, Mukherjee and Kutzelnigg  proposed an elegant
generalisation of normal-ordering w.r.t. such multireference
states \cite{MUKHERJEE:1995:_normalorder, MUKHERJEE:1997:_normalorder, KUT-MUK:1997:_normalorder}.
By examining the form of the above normal-ordering equations when
rotated into an arbitary one-particle basis, they arrived at the
generalised relations
\begin{align}
a^p_s & = \tilde{a}^p_s + \gamma^p_s  \\
a^{pq}_{st} & = \tilde{a}^{pq}_{st} + \gamma^p_s \tilde{a}^q_t
+\gamma^q_t \tilde{a}^p_s -\gamma^p_t \tilde{a}^q_s -\gamma^q_s
\tilde{a}^p_t + \gamma^{pq}_{st} \nonumber\\
& = \tilde{a}^{pq}_{st} + 4 (\gamma^p_s \wedge \tilde{a}^q_t) +
\gamma^{pq}_{st} \label{eq:multitwobodydecomp} \\
a^{pqr}_{stu} &= \tilde{a}^{pqr}_{stu} + \gamma^p_s
\tilde{a}^{qr}_{tu} - \gamma^q_s \tilde{a}^{pr}_{tu}
- \gamma^r_s \tilde{a}^{qp}_{tu} - \gamma^p_t \tilde{a}^{qr}_{su} +
\gamma^q_t \tilde{a}^{pr}_{su}   \nonumber
\\
&- \gamma^r_t \tilde{a}^{pq}_{su}
- \gamma^p_u \tilde{a}^{qr}_{ts}
- \gamma^q_u \tilde{a}^{pr}_{st}
+ \gamma^r_u \tilde{a}^{pq}_{st}
+ \gamma^{pq}_{st} \tilde{a}^r_u
+ \gamma^{pr}_{su} \tilde{a}^q_t\nonumber\\
& + \gamma^{qr}_{tu} \tilde{a}^p_s
-\gamma^{pq}_{su} \tilde{a}^r_t
 - \gamma^{pq}_{ut} \tilde{a}^r_s
 -\gamma^{pr}_{st} \tilde{a}^q_u
 - \gamma^{pr}_{tu} \tilde{a}^q_s
 -\gamma^{qr}_{ts} \tilde{a}^p_u \nonumber \\
& - \gamma^{qr}_{su} \tilde{a}^p_t + \gamma^{pqr}_{stu} \nonumber \\
 &= \tilde{a}^{pqr}_{stu} + 9 (\gamma^p_s \wedge \tilde{a}^{qr}_{tu})
 + 9 (\gamma^{pq}_{st} \wedge \tilde{a}^r_u) + \gamma^{pqr}_{stu}
 \label{eq:multithreebodydecomp}
\end{align}
Let us examine the physical meaning of the above expressions, taking
eqn.\ (\ref{eq:multitwobodydecomp}) as an example. Here, we see that
the original two body operator $a^{pq}_{st}$ is written in terms of
an average over the reference state (the zero-body operator
$\gamma^{pq}_{st}$), a product of a one-body average with a one-body
quasi-particle operator (the terms like $\gamma^p_s \tilde{a}^q_t$), and
a two-body quasi-particle operator $\tilde{a}^{pq}_{st}$.
The quasi-particle operators describe fluctuations about the
reference, because just as in the  usual form of normal ordering, their expectation
values with the reference vanish e.g. $\langle \tilde{a}^p_s\rangle=0,
\langle \tilde{a}^{pq}_{st}\rangle=0$.

\subsection{Application to Canonical Transformation Theory}

The extended normal ordering provides a systematic operator
decomposition which is well suited to Canonical Transformation theory.
At the linearised CTSD level, we wish to decompose the three-body
operators, arising from the commutator $[H, A]$, into lower-body
terms. We can do so by neglecting the effects of the simultaneous
three-body fluctuations described by the operator
$\tilde{a}^{pqr}_{stu}$. For consistency, we should also remove the
fully connected three-body cumulant $\lambda^{pqr}_{stu}$.
First let us rewrite $a^{pqr}_{stu}$ in terms of $a^p_s, a^{pq}_{st}$
be rearranging eqn.\ (\ref{eq:multithreebodydecomp}), and substituting
in the cumulant decomposition of $\gamma^{pqr}_{stu}$
(\ref{eq:threecum}),
we find
\begin{align}
a^{pqr}_{stu}  &= \tilde{a}^{pqr}_{stu} - \gamma^p_s [a^{qr}_{tu} -
\gamma^q_t (a^r_u - \gamma^r_u) + \ldots\nonumber \\& - \gamma^{qr}_{tu}] + \ldots -
\gamma^{pq}_{st} (a^r_u - \gamma^r_u + \ldots \nonumber \\&+ \lambda^{pqr}_{stu} +
\gamma^p_s \lambda^{qr}_{tu} + \ldots \gamma^p_s \gamma^q_r \gamma^t_u
+ \ldots \nonumber\\
&= \tilde{a}^{pqr}_{stu} + 9 (\gamma^p_s \wedge a^{qr}_{tu}) - 36
(\gamma^p_s \wedge \gamma^q_t \wedge a^r_u) + 9(\gamma^{pq}_{st}
\wedge a^r_u)\nonumber \\ 
+&   24 (\gamma^p_s \wedge \gamma^q_t \wedge \gamma^r_u) - 9 (\gamma^{p}_{s} \wedge \gamma^{qr}_{tu}) + \lambda^{pqr}_{stu}
\end{align}
Now dropping $\tilde{a}^{pqr}_{stu}$ and $\lambda^{pqr}_{stu}$ we obtain
the extended normal-ordered decomposition, which we name the {\MK}
decomposition after Mukherjee and Kutzelnigg,
\begin{align}
a^{pqr}_{stu} \Rightarrow& 9 (\gamma^p_s \wedge a^{qr}_{tu}) - 36
(\gamma^p_s \wedge \gamma^q_t \wedge a^r_u) + 9(\gamma^{pq}_{st}
\wedge a^r_u)\nonumber \\ +& 24 (\gamma^p_s \wedge \gamma^q_t \wedge \gamma^r_u)- 9 (\gamma^{pq}_{st} \wedge \gamma^r_u)
\end{align}
Comparing the {\MK} decomposition to our earlier cumulant-type
decomposition (\ref{eq:decomp}) we see that they yield the same
expectation value with the reference function $\Psi_0$ and thus differ
only by terms whose expectation values vanish. In addition to some
different factors, the {\MK} decomposition include additional
operators: a constant term, and the term $\gamma^{pq}_{st} \wedge a^r_u$.
Computationally, both these terms are easily implemented
without affecting the scaling of the original L-CTSD algorithm.

To better understand the differences between the {\MK} and
cumulant-type (CU) decompositions, it is instructive to compare the
two for a simpler example, namely, the decomposition of the
two-particle operator $a^{pq}_{st}$. These are
\begin{align}
a^{pq}_{st} & \Rightarrow 2(\gamma^p_s \wedge a^q_t)&\nonumber \\
&= \frac{1}{2} (\gamma^p_s a^q_t + \gamma^q_t a^p_s -
           \gamma^p_t a^q_s - \gamma^q_s a^p_t) & \mbox{CU} \\
a^{pq}_{st} & \Rightarrow \gamma^p_s (a^q_t - \gamma^q_t) + \gamma^q_t (a^p_s
- \gamma^p_s) \nonumber \\
& - \gamma^p_t (a^q_s - \gamma^q_s) - \gamma^q_s (a^p_t -
\gamma^p_t) + \gamma^{p}_s \gamma^q_t - \gamma^p_t \gamma^q_s & \mbox{\MK}
\end{align}
Here we see that the {\MK} decomposition is expressed in terms of
\textit{fluctuations} e.g.\ $a^q_s - \gamma^q_s$ in the presence of
the field $\gamma^p_t$, while the cumulant decomposition involves the
bare operators $a^q_s$ directly. The neglected term
$\tilde{a}^{pq}_{st}$ in the {\MK} decomposition has the conceptual
meaning of a simultaneous two-particle fluctuation operator, and we
consider this to be conceptually appealing.

Returning to the earlier example that motivated our examination of
alternative operator decompositions, let us now look at the
normal-product decomposition of commutators involving semi-internal
excitation operators, as in eqn.\ (\ref{eq:nonvanishing_term}).
Starting from a single determinantal reference, the extended normal
ordering reduces to the usual normal ordering with respect to a Fermi
vacuum described by eqns (\ref{eq:fermione})-(\ref{eq:threebodyfermidecomp}).
Then, the operator decomposition corresponds to dropping the
three-particle normal-ordered operators $\tilde{a}^{pqr}_{stu}$ in
eqn.\ (\ref{eq:threebodyfermidecomp}). By construction, the remaining
normal-ordered operators e.g.\ $\tilde{a}^{pq}_{rs}$ all have
vanishing expectation value with the Fermi vacuum, and consequently
using the {\MK} decomposition, the expectation values of all
commutators of the form of eqn.\ (\ref{eq:nonvanishing_term}) with
single determinant references \textit{vanish as they should}, in contrast to the cumulant-type decomposition.

Thus we see that the extended normal-ordered MK decomposition offers
some conceptual and formal advantages over our earlier
cumulant-type CU decomposition. Encouraged by these aspects, we have
implemented this decomposition and we now proceed to the numerical
results.

\begin{table}[t]
\begin{center}
\caption{Total energies of FCI and differences of various methods from FCI for the
  simultaneous bond breaking of H$_2$O molecule with CAS$(6e, 5o)$ and cc-pVDZ basis
  sets. The units are \Ehartree. The bond angle is fixed at $<$HOH = 109.57$^\circ$.
  $R_e$ = 0.9929 $\AA$.
  $\tau_s = 10^{-2}$ and $\tau_d = 10^{-2}$ (described in Sec. \ref{sec:orthogonalisation})
  were used in the L-CT calculations.
  See Ref.\ \cite{YAN-CHA:2006:_ctpaper} for the previous L-CT results.}
\label{table:h2o_cc-pVDZ}
\footnotesize
\begin{tabular}{lrrrr}
\hline
\hline
        & $1 R_{e}$ & $2 R_{e}$ & $3 R_{e}$ & $4 R_{e}$ \\
\hline
\hline
FCI     & -76.23885 & -75.94558 & -75.91003 & -75.90872 \\
\hline
RHF     &   0.21718 &   0.37002 &   0.57365 &   0.67159 \\
CASSCF  &   0.16299 &   0.13196 &   0.12302 &   0.12259 \\
CASPT2  &   0.01330 &   0.00843 &   0.00848 &   0.00852 \\
CASPT3  &   0.00377 &   0.00383 &   0.00174 &   0.00158 \\
MR-CI   &   0.00556 &   0.00378 &   0.00296 &   0.00290 \\
MR-CI+Q &  -0.00056 &  -0.00053 &  -0.00066 &  -0.00068 \\
MR-ACPF &   0.00093 &   0.00054 &   0.00020 &   0.00017 \\
MR-AQCC &   0.00231 &   0.00150 &   0.00102 &   0.00098 \\
CCSD    &   0.00384 &   0.02248 &   0.00967 &   0.00200 \\
CCSDT   &   0.00051 &  -0.00238 &  -0.04106 &  -0.04973 \\
L-CTSD(CU)     &   0.00029 &  -0.00097 &  -0.00171 &  -0.00172 \\
L-CTSD(\MK)     &  -0.00077 &  -0.00128 &  -0.00192 &  -0.00192 \\
previous L-CTD(CU)  &   0.00219 &  -0.00056 &   0.00297 &   0.00251 \\
previous L-CTSD(CU) &   0.00061 &  -0.00358 &   0.00301 &   0.00287 \\
\hline
\hline
\end{tabular}
\end{center}
\end{table}

\begin{figure}[b]
\centering
\includegraphics[width=8.0cm]{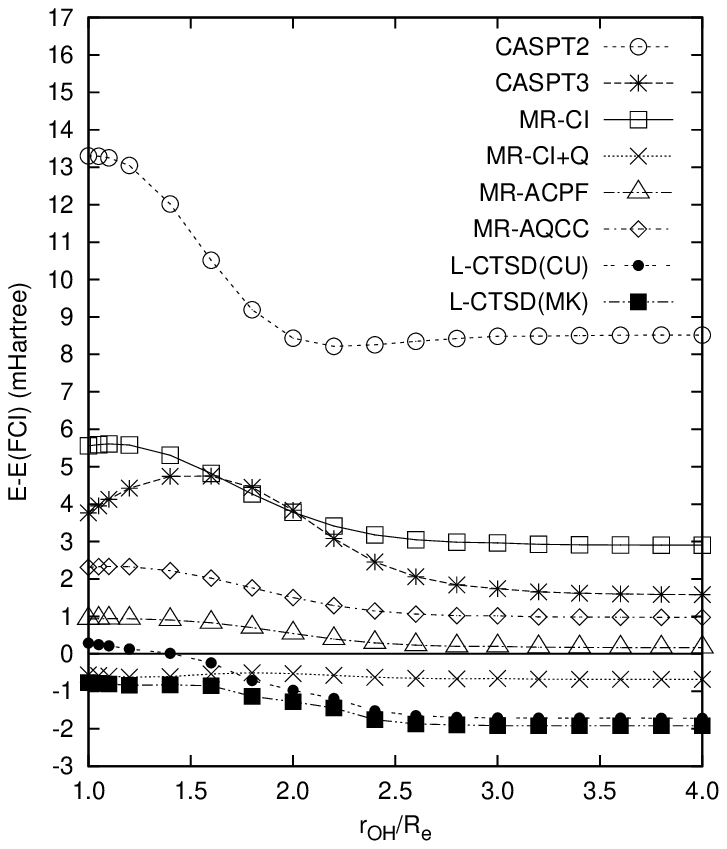}
\caption{Energy differences E-E(FCI) for the
  simultaneous bond breaking of H$_2$O molecule with CAS$(6e, 5o)$ and cc-pVDZ basis
  sets.}
\label{fig:h2o_cc-pVDZ}
\end{figure}

\begin{table}[t]
\begin{center}
\caption{Total energies of MR-CI+Q and differences of various methods from MR-CI+Q for the
  simultaneous bond breaking of H$_2$O molecule with CAS$(6e, 5o)$ and cc-pVTZ basis
  sets. The units are \Ehartree. The bond angle is fixed at $<$HOH = 109.57$^\circ$.
  $R_e$ = 0.9929 $\AA$.
  $\tau_s = 10^{-1}$ and $\tau_d = 10^{-2}$ (described in Sec. \ref{sec:orthogonalisation})
  were used in the L-CT calculations.}
\label{table:h2o_cc-pVTZ}
\begin{tabular}{lrrrr}
\hline
\hline
        & $1 R_{e}$ & $2 R_{e}$ & $3 R_{e}$ & $4 R_{e}$ \\
\hline
\hline
MR-CI+Q & -76.32847 & -76.01591 & -75.97484 & -75.97345 \\
\hline
RHF     &   0.27679 &   0.41697 &   0.60895 &   0.70500 \\
CASSCF  &   0.22228 &   0.18279 &   0.16880 &   0.16815 \\
CASPT2  &   0.01545 &   0.00915 &   0.00904 &   0.00911 \\
CASPT3  &   0.00574 &   0.00646 &   0.00307 &   0.00286 \\
MR-CI   &   0.01005 &   0.00761 &   0.00611 &   0.00603 \\
MR-ACPF &   0.00232 &   0.00176 &   0.00139 &   0.00137 \\
MR-AQCC &   0.00467 &   0.00353 &   0.00280 &   0.00277 \\
CCSD    &   0.00742 &   0.02995 &   0.02724 &   0.01999 \\
CCSDT   &  -0.00055 &  -0.00147 &  -0.03965 &  -0.04866 \\
L-CTSD(CU)  &   0.00214 &   0.00058 &  -0.00081 &  -0.00084 \\
L-CTSD(\MK)  &   0.00186 &  -0.00004 &  -0.00102 &  -0.00103 \\
\hline
\hline
\end{tabular}
\end{center}
\end{table}

\begin{figure}[b]
\centering
\includegraphics[width=8.0cm]{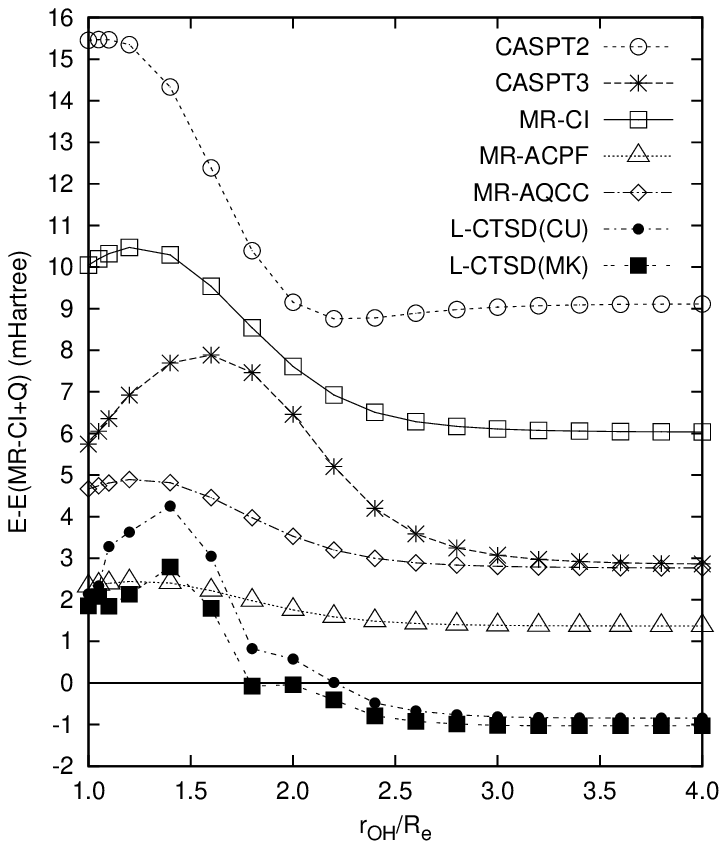}
\caption{Energy differences E-E(MR-CI+Q) for the
  simultaneous bond breaking of H$_2$O molecule with CAS$(6e, 5o)$ and cc-pVTZ basis
  sets.}
\label{fig:h2o_cc-pVTZ}
\end{figure}

\begin{table}[t]
\begin{center}
\caption{Total energies of FCI and differences of various methods from FCI for the
  bond breaking of N$_2$ molecule with CAS$(6e, 6o)$ and 6-31G basis
  sets. The Units are \Ehartree.
  $\tau_s = 10^{-1}$ and $\tau_d = 10^{-2}$ (described in Sec. \ref{sec:orthogonalisation})
  were used in the L-CT calculations.
  See Ref.\ \cite{YAN-CHA:2006:_ctpaper} for the previous L-CT results.}
\label{table:n2_6-31G}
\begin{tabular}{lrrrr}
\hline
\hline
        & $1 \AA$  & $2 \AA$  & $3 \AA$  \\
\hline
\hline
FCI     & -109.04667 & -108.85968 & -108.83905 \\
\hline
RHF     &    0.21143 &    0.55008 &    0.85649 \\
CASSCF  &    0.08551 &    0.08623 &    0.07472 \\
CASPT2  &    0.01372 &    0.00834 &    0.00830 \\
CASPT3  &    0.00558 &    0.00769 &    0.00409 \\
MR-CI   &    0.00268 &    0.00303 &    0.00210 \\
MR-CI+Q &   -0.00012 &   -0.00014 &   -0.00016 \\
MR-ACPF &    0.00092 &    0.00071 &    0.00027 \\
MR-AQCC &    0.00133 &    0.00125 &    0.00069 \\
CCSD    &    0.00685 &   -0.00731 &            \\
CCSDT   &    0.00122 &   -0.05220 &            \\
L-CTSD(CU)     &    0.00142 &   -0.00165 &   -0.00173 \\
L-CTSD(\MK)     &    0.00082 &   -0.00187 &   -0.00250 \\
previous L-CTD(CU)  &    0.00510 &    0.00447 \\
previous L-CTSD(CU) &    0.00646 &    0.00112 \\
\hline
\hline
\end{tabular}
\end{center}
\end{table}

\begin{figure}[b]
\centering
\includegraphics[width=8.0cm]{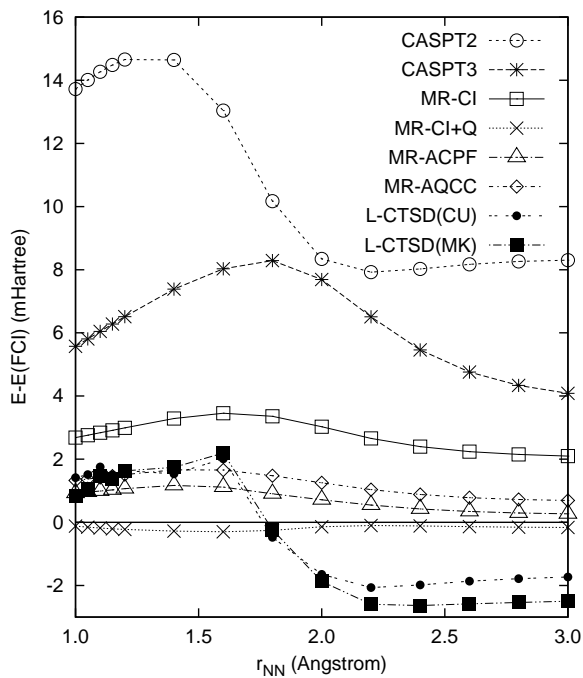}
\caption{Energy differences E-E(FCI) for the bond breaking of N$_2$ molecule with CAS$(6e, 6o)$ and 6-31G basis
  sets.}
\label{fig:n2_6-31G}
\end{figure}

\begin{table}[t]
\begin{center}
\caption{Total energies of MR-CI+Q and differences of various methods from MR-CI+Q for the
  bond breaking of N$_2$ molecule with CAS$(6e, 6o)$ and cc-pVDZ basis
  sets. The units are \Ehartree.
  $\tau_s = 10^{-1}$ and $\tau_d = 10^{-2}$ (described in Sec. \ref{sec:orthogonalisation})
  were used in the L-CT calculations.}
\label{table:n2_cc-pVDZ}
\begin{tabular}{lrrrr}
\hline
\hline
        & $1 \AA$  & $2 \AA$  & $3 \AA$  \\
\hline
\hline
MR-CI+Q & -109.22891 & -108.98376 & -108.96035 \\
\hline
RHF     &    0.29907 &    0.65317 &    0.96627 \\
CASSCF  &    0.18453 &    0.19413 &    0.18316 \\
CASPT2  &    0.02243 &    0.01558 &    0.01616 \\
CASPT3  &    0.00700 &    0.00781 &    0.00375 \\
MR-CI   &    0.00926 &    0.01161 &    0.01011 \\
MR-ACPF &    0.00262 &    0.00245 &    0.00173 \\
MR-AQCC &    0.00419 &    0.00465 &    0.00374 \\
CCSD    &    0.01112 &    0.07424 &            \\
CCSDT   &    0.00177 &   -0.04382 &            \\
L-CTSD(CU)  &    0.00118 &    0.00024 &   -0.00045 \\
L-CTSD(\MK)  &    0.00117 &    0.00162 &    0.00026 \\
\hline
\hline
\end{tabular}
\end{center}
\end{table}

\begin{figure}[b]
\centering
\includegraphics[width=8.0cm]{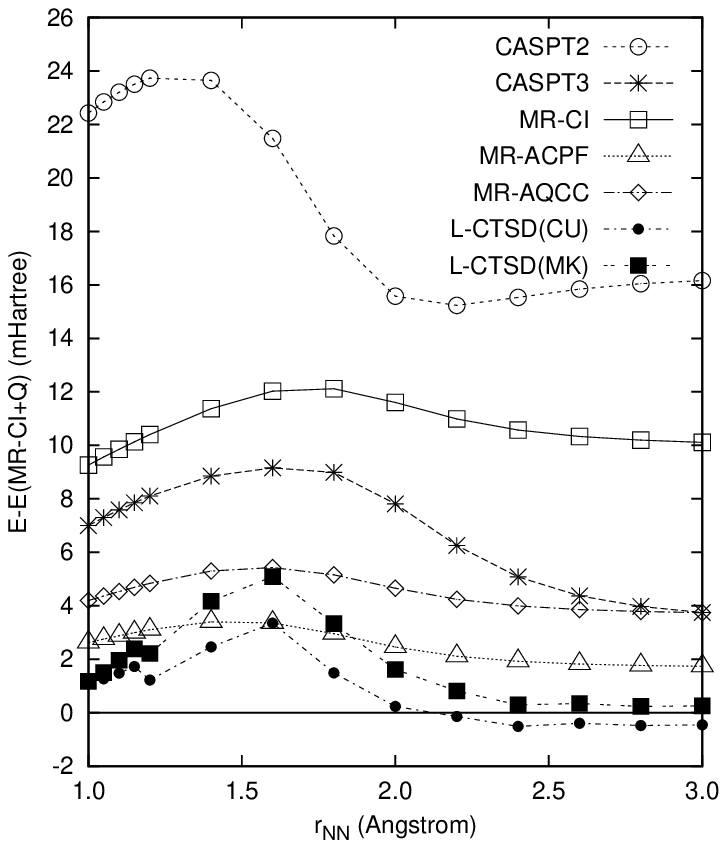}
\caption{Energy differences E-E(MR-CI+Q) for the bond breaking of N$_2$ molecule with CAS$(6e, 6o)$ and cc-pVDZ basis
  sets.}
\label{fig:n2_cc-pVDZ}
\end{figure}

\begin{table}[t]
\begin{center}
\caption{Total energies of MR-CI+Q and differences of various methods from MR-CI+Q for the
  bond breaking of N$_2$ molecule with CAS$(6e, 6o)$ and cc-pVTZ basis
  sets. The units are \Ehartree.
  $\tau_s = 10^{-1}$ and $\tau_d = 10^{-2}$ (described in Sec. \ref{sec:orthogonalisation})
  were used in the L-CT calculations.}
\label{table:n2_cc-pVTZ}
\begin{tabular}{lrrrr}
\hline
\hline
        & $1 \AA$  & $2 \AA$  & $3 \AA$  \\
\hline
\hline
MR-CI+Q & -109.33774 & -109.05871 & -109.03045 \\
\hline
RHF     &    0.36972 &    0.70119 &    1.00458 \\
CASSCF  &    0.25475 &    0.25035 &    0.23571 \\
MR-CI   &    0.01452 &    0.01725 &    0.01516 \\
MR-ACPF &    0.00363 &    0.00324 &    0.00234 \\
MR-AQCC &    0.00625 &    0.00669 &    0.00548 \\
CASPT2  &    0.02399 &    0.01094 &    0.01195 \\
CASPT3  &    0.00753 &    0.01077 &    0.00396 \\
CCSD    &    0.01532 &    0.09593 &            \\
CCSDT   &    0.00021 &   -0.03276 &            \\
L-CTSD(CU)  &    0.00453$^a$&   -0.00006 &   -0.00051 \\
L-CTSD(\MK)  &    0.00249 &    0.00162 &    0.00035 \\
\hline
\hline
\end{tabular}
\begin{flushleft}
a) $\tau_s = 5 \times 10^{-1}$ and $\tau_d = 2 \times 10^{-2}$ were used because of  convergence problems.
\end{flushleft}
\end{center}
\end{table}

\begin{figure}[b]
\centering
\includegraphics[width=8.0cm]{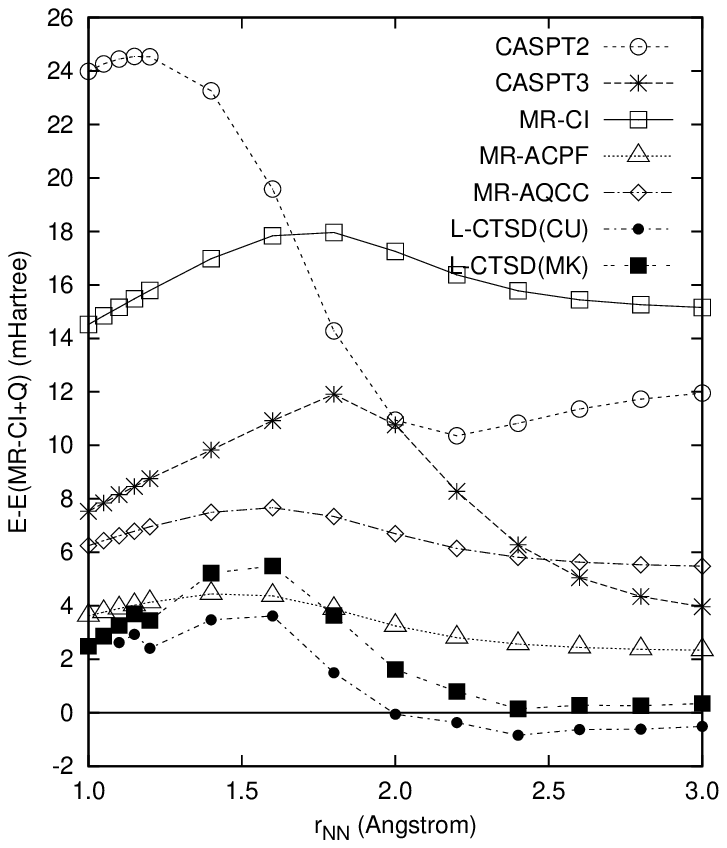}
\caption{Energy differences E-E(MR-CI+Q) for the bond breaking of N$_2$ molecule with CAS$(6e, 6o)$ and cc-pVTZ basis
  sets.}
\label{fig:n2_cc-pVTZ}
\end{figure}

\section{Calculations}
\label{sec:calculations}

\subsection{Water and nitrogen potential energy curves}
\label{sec:water_nitrogen}

We performed prototype multireference CT calculations for the simultaneous bond
breaking curve of the water molecule and the bond breaking curve of the nitrogen
molecule. We chose these molecules to allow a direct comparison with
the results in  our previous paper with the CU decomposition.
\cite{YAN-CHA:2006:_ctpaper}. Here we have used a wider range of basis
sets, including the  cc-pVDZ and
cc-pVTZ basis sets for water and 6-31G, cc-pVDZ, and cc-pVTZ basis sets
for nitrogen \cite{HDP:1972:_631g, DUNNING:1989:_ccpvdz}.
For assessment, we  carried out calculations with  state-of-the-art
internally contracted multireference methods --- second- and
third-order perturbation theory (CASPT2 and CASPT3)
\cite{AMRSW:1990:_caspt2, AMR:1992:_caspt2, WERNER:1996:_caspt3, CEL-WER:2000:_caspt},
configuration interaction (MR-CI) \cite{WER-REI:1982:_icmrci, WER-KNO:1988:_icmrci, KNO-WER:1988:_icmrci},
the {\it a posteriori} size-extensivity corrected configuration
interaction due to Davidson (MR-CI+Q) \cite{LAN-DAV:1974:_mrci_q, DAV-SIL:1977:_mrci_q},
averaged coupled pair functional (MR-ACPF) \cite{GDA-AHL:1988:_mracpf, WER-KNO:1990:_ic_mracpf}
and averaged quadratic coupled-cluster theory (MR-AQCC) \cite{SZA-BAR:1993:_mraqcc}
(both {\it a priori} size-extensivity modifications of configuration
interaction), as well as single-reference coupled cluster calculations
at the CCSD and CCSDT level \cite{BARTLETT:1995:_cc_review, HIRATA:2003:_tce}).
Full configuration interaction (FCI) energies were also used for comparison where available.
The CAS space for the multireference calculations
was six active electrons in five active orbitals [denoted $(6e, 5o)$]
for the water calculations and $(6e, 6o)$ for the nitrogen
calculations. The $1s$ orbitals in O and N atoms were held frozen in
all calculations. For the L-CTSD calculations, we employed both the
cumulant (CU) and normal-ordering (MK) operator decompositions
described in section \ref{sec:ct-theory}.  The internally-contracted
multireference calculations were executed using
{\sc molpro}
\footnote{{\sc molpro}, version 2006.1, a package of ab initio programs, H.-J. Werner, P. J. Knowles, R. Lindh, F. R. Manby, M. Schütz, and others},
the CC calculations using {\sc tce} \cite{HIRATA:2003:_tce}
in {\sc utchem} \cite{UTCHEM:2003}, and the CT calculations
using our own computer program.

Tables \ref{table:h2o_cc-pVDZ}, \ref{table:h2o_cc-pVTZ},
\ref{table:n2_6-31G}, \ref{table:n2_cc-pVDZ}, and
\ref{table:n2_cc-pVTZ} present the errors in the total energies of
various methods as measured from FCI or (in the larger basis sets)
MR-CI+Q at several points across the potential curve.  These errors
are plotted in Figures \ref{fig:h2o_cc-pVDZ}, \ref{fig:h2o_cc-pVTZ},
\ref{fig:n2_6-31G}, \ref{fig:n2_cc-pVDZ}, and \ref{fig:n2_cc-pVTZ}.


Comparing all the different methods, in the calculations where FCI
energies were available, MR-CI+Q provided the smallest maximum
absolute error (MAE) and non-parallelity error (NPE) and for this
reason was used as the benchmark method when FCI energies could not be
obtained. The general order of accuracy in terms of MAE from most to
least accurate was MR-CI+Q $\approx$ MR-ACPF $\approx$ L-CTSD(CU),
L-CTSD(MK) $\approx$ MR-AQCC $>$ CASPT3 $\approx$ MR-CI $>$ CASPT2.
While the MAE of L-CTSD(CU) and L-CTSD(MK) was comparable to that of
MR-ACPF and MR-AQCC, the NPE was larger; in the intermediate region
the shapes of the curves somewhat resembled the CASPT3 curve. In the
equilibrium region, the L-CTSD energies were similar in accuracy to
CCSDT.

\begin{table}
\begin{center}
\caption{Maximum absolute error (MAE) and non-parallelity error (NPE) of L-CTSD(CU) and L-CTSD(\MK). The units are m\Ehartree.}
\label{table:mae_npe_cu_no}
\begin{tabular}{p{3cm}ccccc}
\hline
\hline
               & \multicolumn{2}{c}{L-CTSD(CU)}  & & \multicolumn{2}{c}{L-CTSD(\MK)} \\
\hline
               & MAE  & NPE  & & MAE  & NPE  \\
\hline
\hline
H$_2$O/cc-pVDZ & 1.72 & 2.01 & & 1.92 & 1.11 \\
H$_2$O/cc-pVTZ & 4.25 & 5.10 & & 2.79 & 3.82 \\
N$_2$/6-31G    & 2.07 & 4.08 & & 2.54 & 4.74 \\
N$_2$/cc-pVDZ  & 3.35 & 3.83 & & 5.09 & 4.86 \\
N$_2$/cc-pVTZ  & 3.62 & 4.46 & & 5.49 & 5.35 \\
\hline
\hline
\end{tabular}
\end{center}
\end{table}

\begin{table}
\begin{center}
\caption{Spectroscopic constants for N$_2$ molecule by various methods
  with 6-31G, cc-pVDZ, and cc-pVTZ basis sets.
The dissociation energy $D_e$ was obtained with additional  atomic  calculations
for the nitrogen atom.
}
\label{table:spec_n2}
\begin{tabular}{p{3.0cm}rrr}
\hline
\hline
        & $R_{e}$  & $\omega_{e}$  & $D_{e}$    \\
& $\AA$    & cm$^{-1}$  & kcal/mol \\
\hline
\hline
6-31G \\
\hline
FCI         &  1.134 86 & 2208.27 & 168.45 \\
\hline
RHF         & -0.045 74 &  452.94 & -103.25 \\ 
CCSD        & -0.006 77 &   77.53 &   -6.27 \\ 
CCSDT       & -0.002 06 &   26.93 &   -1.38 \\ 
CASSCF      & -0.003 77 &   29.35 &   -8.87 \\ 
CASPT2      & -0.000 91 &    3.71 &   -3.85 \\ 
CASPT3      & -0.001 01 &    8.80 &   -1.62 \\ 
MR-CI       & -0.000 33 &    3.14 &   -1.24 \\ 
MR-CI+Q     &  0.000 10 &   -0.15 &   -0.02 \\ 
MR-ACPF     & -0.000 16 &    1.68 &   -0.49 \\ 
MR-AQCC     & -0.000 20 &    1.99 &   -0.55 \\ 
L-CTSD(CU)  &  0.001 11 &  -15.35 &   -0.67 \\ 
L-CTSD(\MK) &  0.000 53 &  -10.34 &   -1.21 \\ 
\hline
\hline
cc-pVDZ \\
\hline
MR-CI+Q               &  1.120 36 & 2321.25 & 200.59 \\
\hline
RHF         & -0.043 06 &  436.76 &  -88.43 \\ 
CCSD        & -0.007 54 &   87.11 &   -8.62 \\ 
CCSDT       & -0.001 87 &   24.90 &   -1.79 \\ 
CASSCF      & -0.005 89 &   43.95 &   -3.58 \\ 
CASPT2      & -0.001 21 &    4.48 &   -4.11 \\ 
CASPT3      & -0.001 07 &    7.77 &   -2.40 \\ 
MR-CI       & -0.001 10 &    8.71 &   -2.83 \\ 
MR-ACPF     & -0.000 47 &    3.34 &   -0.48 \\ 
MR-AQCC     & -0.000 62 &    4.39 &   -0.48 \\ 
L-CTSD(CU)  & -0.001 25 &   -4.77 &   -0.42 \\ 
L-CTSD(\MK) & -0.002 07 &   -6.20 &   -1.13 \\ 
\hline
\hline
cc-pVTZ \\
\hline
MR-CI+Q               &  1.104 76 & 2332.40 & 216.00 \\
\hline
RHF         & -0.037 59 &  398.68 &  -95.61 \\ 
CCSD        & -0.008 05 &   90.91 &   -8.30 \\ 
CCSDT       & -0.001 65 &   22.60 &   -0.15 \\ 
CASSCF      & -0.001 10 &   18.63 &  -12.24 \\ 
CASPT2      & -0.000 45 &   -4.75 &   -6.99 \\ 
CASPT3      & -0.001 23 &   10.06 &   -2.41 \\ 
MR-CI       & -0.001 24 &   10.58 &   -4.22 \\ 
MR-ACPF     & -0.000 43 &    3.09 &   -0.37 \\ 
MR-AQCC     & -0.000 63 &    4.87 &   -0.43 \\ 
L-CTSD(\MK) & -0.002 49 &    2.02 &   -1.68 \\ 
\hline
exptl                 &  1.107 68 & 2358.57 & 228.4  \\
\hline
\hline
\end{tabular}
\end{center}
\end{table}


The MAE and NPE for the two CT operator decompositions L-CTSD(CU) and L-CTSD(\MK) are compared in Table
\ref{table:mae_npe_cu_no}.
We find that the two operator decompositions performed quite similarly in
these systems, although the MAE of L-CTSD(CU) was  slightly smaller.
For comparison, we have also included the L-CTSD(CU) energies  from
our calculations in our earlier work \cite{YAN-CHA:2006:_ctpaper}. We note that our
new L-CTSD(CU) energies are \textit{significantly improved}, particularly in the intermediate
dissociation region. This is a result of the new numerical algorithm,
described in \ref{sec:orthogonalisation}, which allowed us to significantly reduce the truncation
of the operator manifold that we used in our previous work.
However, the curves of the new L-CTSD in the figures are not  completely
smooth due to some remaining operator truncation effects in the
numerical solution and removal of
this non-smooth behaviour will be addressed in future work.



Table \ref{table:spec_n2} shows the spectroscopic constants of N$_2$
computed by fitting the potential curves.
Compared to the available FCI results in
the 6-31G basis, MR-CI+Q once again came closest for all
spectroscopic parameters ($R_e$, $\omega_e$, $D_e$) while the related
 MR-ACPF and MR-AQCC methods behaved very similarly to
 MR-CI+Q. Comparing CT against  the other methods,  different trends were observed
for different quantities. For the dissociation energies, we found that 
MR-CI+Q $>$ MR-ACPF $\approx$ MR-AQCC $\approx$ L-CTSD(CU) $>$
L-CTSD(MK) $\approx$ CCSDT $>$ CASPT3 $>$ CASPT2 $>$ CCSD.
For frequencies, in the cc-pVDZ and cc-pVTZ basis L-CTSD was
comparable in accuracy to MR-ACPF/MR-AQCC (though with errors in the
opposite direction) and better than
than those of CCSDT, while the equilibrium bond distances were less
accurate than MR-ACPF/MR-AQCC though still comparable to
CCSDT. L-CTSD(CU) and L-CTSD(\MK) generally  performed similarly, although the spectroscopic constants for
 L-CTSD(CU) with cc-pVTZ could not be obtained because of convergence
 problems at the fitting geometries. The small non-smoothness in the
 potential energy curves resulting from the numerical approximations  in solving the
CT equations may also be a factor in  the less systematic
 errors of the CT methods for $R_e$ and $\omega_e$.


Thus to summarise, the overall performance of L-CTSD(CU) and L-CTSD(MK) for these
potential energy curves was   competitive with the best
multireference methods, such as MR-ACPF and MR-AQCC, particularly for energetic
 quantities such as the MAE and $D_e$. The shapes of the curves in the
 intermediate regions looked somewhat like the CASPT3 curves, though
 with significantly smaller absolute errors. The  spectroscopic constants
  $\omega_e, R_e$  and the non-parallelity error from L-CTSD  were slightly
 less accurate than from MR-ACPF and MR-AQCC and  this  was in part related
 to our numerical  approximations in solving the CT equations.





\subsection{Size-consistency}

\begin{table}[t]
\begin{center}
\caption{Size consistency errors (m\Ehartree) of CISD, ACPF, AQCC, CCSD, L-CTSD(CU), and L-CTSD(MK) calculations.}
\label{table:sizec_errors}
\begin{tabular}{p{2.5cm}rrrrrr}
\hline
\hline
           & Be + He    & Be + 2He   & Be + 3He   & Be + 4He \\
\hline
CISD       &  3.10   &  6.23   &  9.46   & 12.83 \\
ACPF       & -0.56   & -0.76   & -0.86   & -0.92 \\
AQCC       &  1.98   &  2.38   &  2.66   &  2.91 \\
CCSD       &  0.00   &  0.00   &  0.00   &  0.00 \\
L-CTSD(CU) &  0.00   &  0.00   &  0.00   &  0.00 \\
L-CTSD(MK) &  0.00   &  0.00   &  0.00   &  0.00 \\
\hline
\hline
           & N$_2$ + He & N$_2$ + 2He& N$_2$ + 3He& N$_2$ + 4He \\
\hline
CISD       &  1.96   &  4.00   &  6.12   &  8.31 \\
ACPF       & -0.41   & -0.71   & -0.94   & -1.11 \\
AQCC       & -0.57   & -0.98   & -1.28   & -1.50 \\
CCSD       &  0.00   &  0.00   &  0.00   &  0.00 \\
L-CTSD(CU) &  0.00   &  0.00   &  0.00   &  0.00 \\
L-CTSD(MK) &  0.00   &  0.00   &  0.00   &  0.00 \\
\hline
\hline
\end{tabular}
\end{center}
\end{table}

\begin{table*}[t]
\begin{center}
\caption{Energy difference of all-electron and frozen-core atomic calculations, i.e. $E(\mathrm{all\,\,electron}) - E(\mathrm{frozen\,\,core})$, by various methods with 6-31G basis sets.}
\label{table:sizec_atom_diff}
\begin{tabular}{p{2.5cm}rrrrrr}
\hline
\hline
            & Be        & Ne         & Mg         & Ar         & Ca         \\
\hline
FCI         & -0.00081 & -0.00078 & -0.00276 & -0.00195 & -0.00322 \\
CISD        & -0.00075 & -0.00077 & -0.00252 & -0.00187 & -0.00292 \\
CCSD(T)     & -0.00080 & -0.00078 & -0.00277 & -0.00196 & -0.00322 \\
CCSD        & -0.00078 & -0.00077 & -0.00265 & -0.00189 & -0.00308 \\
AQCC        & -0.00156 & -0.00098 & -0.00452 & -0.00205 & -0.00505 \\
ACPF        & -0.00335 & -0.00090 & -0.00503 & -0.00199 & -0.00535 \\
MP3         & -0.00099 & -0.00076 & -0.00275 & -0.00185 & -0.00324 \\
MP2         & -0.00105 & -0.00091 & -0.00289 & -0.00219 & -0.00318 \\
L-CTSD(CU)  &  0.05377 &  0.11300 &  0.03612 &  0.03801 &  0.02855 \\
L-CTSD(\MK) &  0.00004 &  0.00460 & -0.00289 &  0.00660 & -0.00320 \\
\hline
\hline
\end{tabular}
\end{center}
\end{table*}

As is well recognized,  size-consistency is a crucial
requirement for any correlation model  to obtain
chemically accurate results in systems with many correlated electrons.
As discussed in our initial  work \cite{YAN-CHA:2006:_ctpaper}, the L-CT theory is naturally
size-consistent. One way to see this is to observe that the energy is obtained as the
expectation value of an effective Hamiltonian that contains only
connected contributions by virtue of its construction  via
a commutator expansion (\ref{eq:decompbch}). Here we verify the size
consistency property of L-CT theory through explicit numerical calculations on
 supermolecules. We have chosen to use supermolecules that contain  more than one type of molecule  since certain approximate size-extensive
 theories such as the ACPF and AQCC methods (which   modify
the non-size-consistent CISD method) are rigorously size-consistent only in the
special case when the supermolecule is made of $n$ noninteracting
{\it identical} subsystems.



Table \ref{table:sizec_errors} gives the size
consistency errors of L-CTSD, CISD, CCSD, ACPF and AQCC calculations
for the Be + $n$ He and N$_2$ + $n$
 He, respectively. All calculations used the HF wavefunction as  the
 reference and the molecules/atoms were each separated by a
 distance of 1000 bohr.
Size consistency implies the condition $E(\mathrm{A} + n \mathrm{B}) =
 E(\mathrm{A}) + n E(\mathrm{B})$. As can be seen, the L-CTSD and CCSD
 calculations are  rigorously size-consistent while those of CISD,
 ACPF and AQCC steadily increase.

Consider now the ACPF energy functional of the noninteracting system $\mathrm{A} + n \mathrm{B}$,  given by
\begin{align}
F^\mathrm{ACPF}_{\mathrm{A}+n\mathrm{B}} = \frac{\langle H_\mathrm{A} \rangle + n \langle H_\mathrm{B} \rangle}
                                                {1 + (2/N) \langle \delta_\mathrm{A} | \delta_\mathrm{A} \rangle + (2n / N)  \langle \delta_\mathrm{B} | \delta_\mathrm{B} \rangle} \label{acpf_functional}
\end{align}
where $\langle H_\mathrm{A} \rangle = \langle \Psi_\mathrm{A}
|H_\mathrm{A}| \Psi_\mathrm{A} \rangle$, $N$ is the total number of electrons, which is equal to $N_A + n
N_B$, and $\delta$ denotes the orthogonal correlation component of
$\Psi$, e.g. $\Psi_\mathrm{A} = \Psi_{0\mathrm{A}} + \delta_\mathrm{A}$.
If $A = B$ in eqn.\ (\ref{acpf_functional}), we  readily
confirm that $F^\mathrm{ACPF}_{(1+n)\mathrm{B}} = (1+n)
F^\mathrm{ACPF}_{\mathrm{B}}$ and the energy is size-consistent. The size consistency error in the functional is obtained as,
{\footnotesize
\begin{align}
\epsilon(n) & = F^\mathrm{ACPF}_{\mathrm{A}+n\mathrm{B}} - F^\mathrm{ACPF}_\mathrm{A} - n \, F^\mathrm{ACPF}_\mathrm{B} \nonumber \\
            & =
\frac{ R_\mathrm{A} R_\mathrm{B} ( N_\mathrm{A} \langle H_\mathrm{B} \rangle + N_\mathrm{B} \langle H_\mathrm{A} \rangle )
                 -R_\mathrm{A}^2 N_\mathrm{B} \langle H_\mathrm{B} \rangle
                 -R_\mathrm{B}^2 N_\mathrm{A} \langle H_\mathrm{A} \rangle
         }
     { R_\mathrm{A} R_\mathrm{B}^2 + R_\mathrm{A}^2 R_\mathrm{B} / n  }
\end{align}
}where $R_\mathrm{A} = N_\mathrm{A} + 2 \langle \delta_\mathrm{A} |
\delta_\mathrm{A} \rangle $. The errors of ACPF and AQCC indeed appear to behave as the above function.  (Note
that $\epsilon(n=\infty)$ does not vanish).


\subsection{Density dependence}

Rather than considering the scaling behaviour of the energy as we
increase the number of molecules, we can also consider
the complementary trend of  going to atoms
with larger and larger nuclear charge $Z$ (and consequently more and
more electrons in the same region of space).
In essence, this measures the density dependence of the energy.
To study the
behaviour of the CT and other methods under this condition, we
chose five closed-shell atoms, two rare gas atoms (Ne[10e] and
Ar[18e]) and three alkaline earth metals (Be[4e], Mg[12e] and
Ca[20e]).

\begin{figure}[h]
\centering
\includegraphics[width=10.0cm]{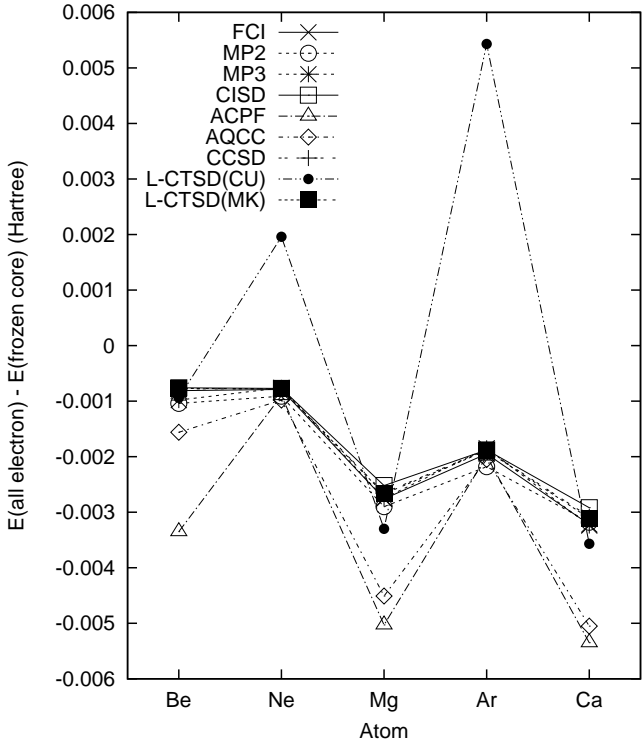}
\caption{Density scaling: Energy difference $E(\mathrm{all\,\,electron}) - E(\mathrm{frozen\,\,core})$ for atomic calculations shown in Tables \ref{table:sizec_atom_diff}.}
\label{fig:sizec_atom}
\end{figure}

Table \ref{table:sizec_atom_diff} and figure \ref{fig:sizec_atom}
present the core-correlation energies, defined as the energy difference
between all-electron and frozen-core atomic calculations, using 6-31G basis
sets. The frozen-core calculations correlate eight and two electrons
in the valence orbitals for the rare gas atoms and alkaline earth
metals, respectively, and represent 98.3\% (Be), 99.3\% (Ne), 92.1\% (Mg),
95.3\% (Ar), and 88.7\% (Ca) of the all-electron correlation energies.
The  energy difference between the all-electron and frozen-core
calculations  is the core correlation energy from the core-valence and core-external excitations. Since we can regard
valence-electron correlation as a size-intensive quantity that
is described by a fixed, i.e. O(1), small number of valence electrons at
a roughly constant valence electron density \footnote{This separation
  of core and valence densities can be made precise in Thomas-Fermi
  theory, see e.g. E. H. Lieb, Rev. Mod. Phys. {\bf 48}, 553 (1976).}, the rest of the
correlation for the bulk of the electrons, i.e. the core correlation,
must contain the main density dependence as we change the number of electrons and
nuclear charge $Z$.


 Compared to the exact  FCI
core correlation energies, it is clear that ACPF, AQCC and L-CTSD(CU)
have difficulty  reproducing the correct behaviour. In particular large errors are found in the ACPF
and AQCC calculations for the alkaline earth metals and in the L-CTSD(CU)
calculations of the rare gas atoms. By contrast, the size-inconsistent
CISD method as well as the MP2 and MP3 methods are able to capture the correct behaviour of  the core
correlation. This illustrates the difficulty in finding an ad-hoc
size-consistency correction, as employed in ACPF and AQCC, that works
under all conditions.
Most interestingly, the new operator decomposition  in
L-CTSD(MK) behaves  much better than L-CTSD(CU) and reproduces the
correct behaviour.

\subsection{FeO binding curve}

\begin{table*}[t]
\begin{center}
\caption{Total energies of MR-CI+Q (\Ehartree) and differences (m\Ehartree) of various methods from MR-CI+Q for the ground $1^5\Delta$ state of FeO molecule.}
\label{table:feo_diff}
\begin{tabular}{p{2.5cm}rrrrr}
\hline
\hline
                         &   1.50$\AA$ &   1.57$\AA$ &   1.65$\AA$ &   1.72$\AA$ &   2.00$\AA$ \\
\hline
MRCI+Q                   & -1337.65843 & -1337.67007 & -1337.67302 & -1337.66923 & -1337.62980 \\
\hline
CASSCF                   &      299.22 &      294.87 &      290.00 &      284.82 &      265.74 \\
CASPT2                   &       -8.20 &       -7.14 &       -5.73 &       -4.12 &        0.57 \\
CASPT3                   &       41.79 &       41.78 &       40.57 &       38.46 &       28.73 \\
MRCI                     &       21.78 &       21.41 &       20.93 &       20.37 &       17.92 \\
MRACPF                   &        0.51 &        0.40 &        0.32 &        0.29 &        0.45 \\
MRAQCC                   &        4.61 &        4.46 &        4.30 &        4.16 &        3.79 \\
L-CTSD(CU)$^\mathrm{a}$  &           c &           c &        9.17 &        9.43 &        7.67 \\
L-CTSD(CU)$^\mathrm{b}$  &           c &           c &        6.85 &        6.85 &        3.51 \\
L-CTSD(\MK)$^\mathrm{a}$ &        3.09 &        2.73 &        3.00 &        3.85 &        3.52 \\
L-CTSD(\MK)$^\mathrm{b}$ &        0.83 &        1.21 &        1.66 &        2.25 &       -0.16 \\
\hline
\hline
\end{tabular}
\begin{flushleft}

a) $\tau_s = 3.0 \times 10^{-1}$ and $\tau_d = 5.0 \times 10^{-2}$.

b) $\tau_s = 1.5 \times 10^{-1}$ and $\tau_d = 5.0 \times 10^{-2}$.

c) Not converged.
\end{flushleft}
\end{center}
\end{table*}

\begin{table}[b]
\begin{center}
\caption{Spectroscopic constants for the ground $1^5\Delta$ state of FeO molecule.}
\label{table:feo_constant}
\begin{tabular}{p{2.5cm}rrrrr}
\hline
\hline
                         & $R_e (\AA)$ & $\omega_e (\mathrm{cm}^{-1})$ \\
\hline
CASSCF                   & 1.703 & 691.1 \\
CASPT2                   & 1.620 & 913.8 \\
CASPT3                   & 1.657 & 755.2 \\
MR-CI                    & 1.641 & 844.4 \\
MR-CI+Q                  & 1.635 & 863.2 \\
MR-ACPF                  & 1.636 & 863.5 \\
MR-AQCC                  & 1.637 & 858.7 \\
L-CTSD(\MK)$^\mathrm{a}$ & 1.631 & 914.0 \\
L-CTSD(\MK)$^\mathrm{b}$ & 1.630 & 911.1 \\
\hline
exptl.              & 1.616 & 880   \\
\hline
\hline
\end{tabular}
\begin{flushleft}
a) $\tau_s = 3.0 \times 10^{-1}$ and $\tau_d = 5.0 \times 10^{-2}$.

b) $\tau_s = 1.5 \times 10^{-1}$ and $\tau_d = 5.0 \times 10^{-2}$.

\end{flushleft}
\end{center}
\end{table}

\begin{figure*}[t]
\centering
\includegraphics[width=17.0cm]{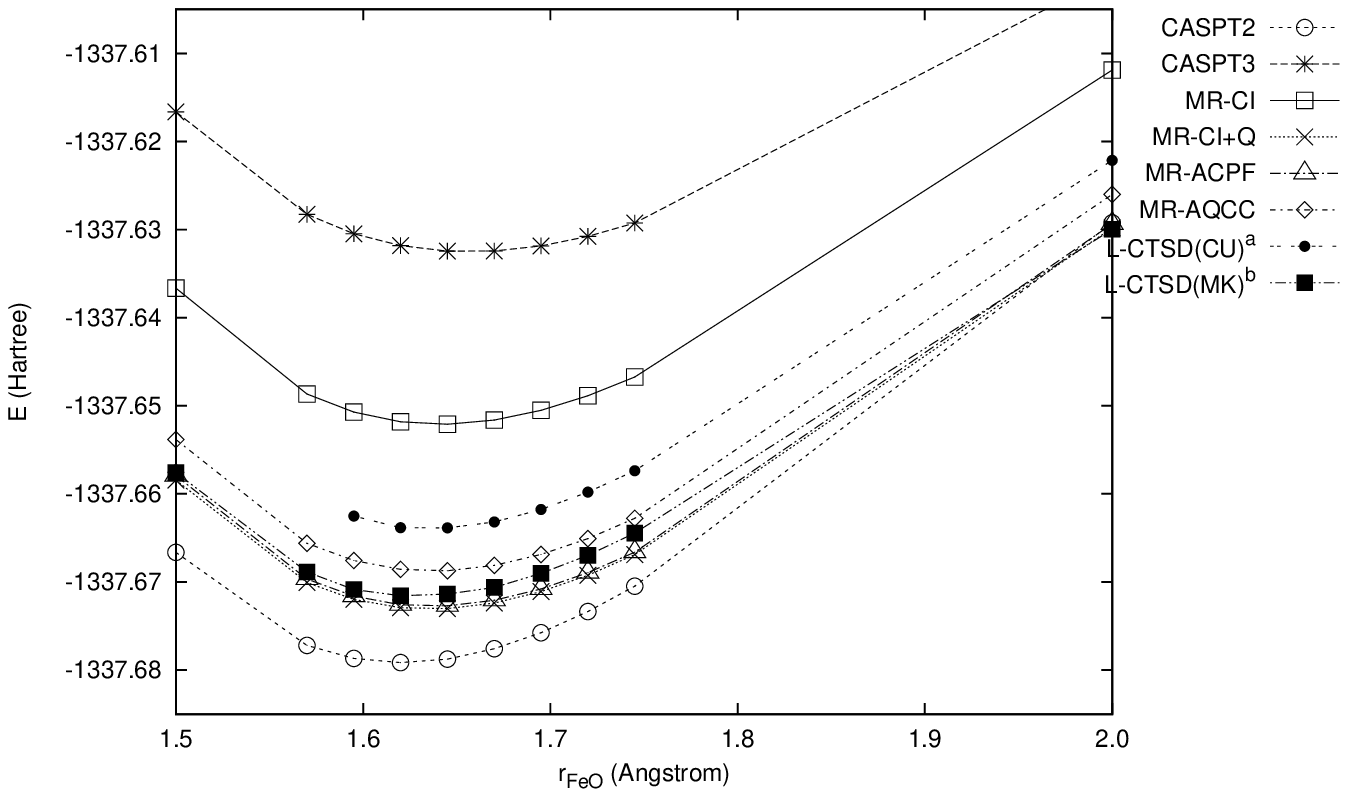}
\caption{Potential curve for the ground $1^5\Delta$ state of FeO molecule. a) $\tau_s = 3.0 \times 10^{-1}$ and $\tau_d = 5.0 \times 10^{-2}$. b) $\tau_s = 1.5 \times 10^{-1}$ and $\tau_d = 5.0 \times 10^{-2}$.}
\label{fig:feo_curve}
\end{figure*}

As a realistic example of a  difficult multireference problem, we
calculated the potential curve for the ground $1^5\Delta$ state of the
FeO molecule.
ANO basis sets \cite{ROOSANO:1990:_roos_ano_H_Ne, ROOSANO:1995:_roos_ano_Sc_Zn} of DZP quality were used,
$[21s15p10d6f]/(5s4p3d1f)$ and $[14s9p4d]/(3s2p1d)$ for the Fe and O
basis, respectively.  To facilitate the setup for the multireference
calculations, the initial orbitals were obtained from  closed-shell
RHF calculations for the $1^1\Sigma$ state. The ten lowest
lying orbitals for 20 electrons
\begin{equation}
(1\sigma)^2(2\sigma)^2(3\sigma)^2(4\sigma)^2(5\sigma)^2(6\sigma)^2(1\pi)^4(2\pi)^4
\end{equation}
were held frozen for the CASSCF and subsequent dynamic correlation
calculations.  We verified that the errors made by this orbital
restriction were almost constant within 1 m$E_h$ and thus would not
affect the shapes of the potential curves. The orbital $(7\sigma)^2$
was treated as an external core orbital, which was optimized by
CASSCF and then correlated. The remaining 12 electrons were fully correlated with 12 active orbitals
\begin{equation}
(8\sigma)^2(9\sigma)^1(10\sigma)^0(11\sigma)^0(3\pi)^4(4\pi)^2(5\pi)^0(1\delta)^3
\end{equation}
(the occupations are based on the ROHF configuration of the $^5 \Delta$ state) for CAS denoted as
$(12e, 12o)$. This CAS is derived from Fe $3d$ and $4s$ orbitals,
oxygen $2p$ orbitals, and the third bonding and antibonding $\pi$ orbitals, which are
formed from the oxygen $2p_\sigma$ orbital mixing with some Fe
$4p_\sigma$ \cite{BLK:1990:_feo_nih_tio, HARRISON:2000:_transition_metal}.

Figure \ref{fig:feo_curve} shows the potential curves of FeO computed
by various multireference methods. As exact energies are not available
for this system, we report the differences from MR-CI+Q
energies  in Table \ref{table:feo_diff}.
Clearly, the MR-ACPF, MR-AQCC, and L-CTSD(MK) curves  are all very close to
each other, while the MR-CI and CASPT2/CASPT3 curves are significantly
further away.
The MR-ACPF curve follows the MR-CI+Q curve with deviations of
less than 0.5 m\Ehartree, while the MR-AQCC curve is also nearly parallel
with  deviations of 3.8-4.6 m\Ehartree. The L-CTSD(CU) and L-CTSD(MK)
curves were shifted relative to each other; the L-CTSD(MK) energies
were significantly closer to the MRCI+Q energies, with deviations of less than 2.3 m\Ehartree.
CASPT2 seemed to overestimate the correlation energy, while going to
the third order CASPT3 over-corrected too much in the opposite
direction and strongly underestimated the correlation energy.




Table \ref{table:feo_constant} shows the spectroscopic constants
measured from the potential curves.
While the basis used is probably too small for direct comparison
to experiment,  we see that in relation to the experimental results,
MR-CI and related modifications MR-CI+Q,  MR-ACPF and MR-AQCC give frequencies
that are too low and bond-lengths that are too long, while L-CTSD(MK)
gives  frequencies that are too high and slightly improved
bond-lengths. As we have already seen in the difference between
the CASPT2 and CASPT3   curves, multireference perturbation theory seemed to break down in this molecule.


\subsection{Timings}

It is our intention that the CT theory should be practically
applicable to problems of reasonable size, and let us now examine the
computational timings for the multireference calculations on the FeO
molecule we have just discussed. These are shown in Table
\ref{table:timings}. All timings were obtained on a single CPU of the
Altex system (Itanium 1.5GHz) at the Research Center for Computational Science, Okazaki.  
As can be seen, the MR-CI based methods were two to three orders of magnitude
more expensive than CASPT2. L-CTSD(MK) displayed very satisfactory
performance. Even in our protoypte CT implementation, which did not
use point-group symmetry, the single-point energy calculation
\textit{took less time than even the CASPT2 calculation, while
  providing a significantly better accuracy competitive with MR-ACPF}.


\begin{table}[bh]
\begin{center}
\caption{Timings for different multireference methods for a single
  point calculation on the FeO curve. Note that the L-CTSD calculation
  did not use point-group symmetry, while $C_{2v}$ symmetry was used
  in all the other calculations. The time for the CASSCF calculation
  is not included.}
\label{table:timings}
\begin{tabular}{p{2.5cm}rrrrr}
\hline
\hline
                         & Time/s \\
\hline
CASPT2                   &   5900 \\
CASPT3                   &  17000 \\
MR-CI+Q                  & 158000 \\
MR-ACPF                  & 168000 \\
L-CTSD(\MK)$^\mathrm{a}$ &   4500 \\
\hline
\hline
\end{tabular}
\begin{flushleft}
a) The time for constructing density matrices is not included.
\end{flushleft}
\end{center}
\end{table}

\section{Summary and Conclusions}

We have been developing the Canonical Transformation theory to describe dynamic
correlation in multireference problems. The theory  uses a size-extensive
unitary exponential acting on a multireference  function.
In our initial work, we introduced a central approximation that
rendered the  manipulation of this ansatz practical, namely a
cumulant-based operator decomposition.
This choice of decomposition is not unique, however, and in the
current work we introduced a new operator decomposition,
based on the extended normal ordering of Mukherjee and Kutzelnigg \cite{MUKHERJEE:1995:_normalorder, MUKHERJEE:1997:_normalorder, KUT-MUK:1997:_normalorder},
which possesses attractive formal and conceptual features.

We carried out calculations at the Linearised Canononical
Transformation Theory Singles and Doubles (L-CTSD) level using both
our earlier cumulant-based and current Mukherjee-Kutzelnigg operator
decompositions.  In studies of the water, nitrogen, and iron-oxide
binding curves, we found the accuracy of L-CTSD to be competitive
with some of the best existing multireference methods such as the
Multireference Averaged Coupled Pair Functional, while the
computational cost was two-three orders of magnitude less and
comparable to that of Complete-Active-Space Second Order Perturbation
Theory. Compared to  our  earlier work, our results and
computational timings were greatly improved, in part due  to the use
of a new numerical algorithm for converging the Canonical
Transformation equations.

\section{Acknowledgements}

This work was supported by Cornell University, the National Science
Foundation CAREER program CHE-0645380, and the David and Lucile
Packard Foundation. We also acknowledge a grant of computer time at the Research
Center for Computational Science, Okazaki, Japan, with which some of
these calculations were performed.

\section{Appendix: Implementing Canonical Transformation Theory}

\subsection{Recapitulation}

In our previous implementation of the CT algorithm \cite{YAN-CHA:2006:_ctpaper} we solved the residual
equations using the following skeletal algorithm
\begin{enumerate}
\item Set up the electronic Hamiltonian $H$ and the one- and
two-particle density matrices of a reference wavefunction.
\item Compute the transformed Hamiltonian $\bar{H}_{1,2}$ (eqn.\ (\ref{eq:decompbch})).
\item Compute the residuals of CT amplitude equations.
\begin{align}
R^{p}_{s} &= \langle [\bar{H}_{1,2}, a^{p}_{s} - a^{s}_{p}]_{1,2} \rangle \label{eq:residual_a1}\\
R^{pq}_{st} &= \langle [\bar{H}_{1,2}, a^{pq}_{st} - a^{st}_{pq}]_{1,2} \rangle  \label{eq:residual_a2}
\end{align}
\item Update the amplitudes by adding the preconditioned residuals
\begin{align}
A^{p}_{s} &\leftarrow A^{p}_{s} - R^{p}_{s} / D^{p}_{s}
\label{eq:diag1} \\
A^{pq}_{st} &\leftarrow A^{pq}_{st} - R^{pq}_{st} / D^{pq}_{st} \label{eq:diag2}
\end{align}
where the factors $1/D^{p}_{s}$ and $1/D^{pq}_{st}$ are the diagonal preconditioners.
\item Repeat (2)-(4) until convergence.
\end{enumerate}
In addition, we employed a somewhat
complicated division of the optimisation process into different steps
involving different classes of excitations in the $A$ operator.

\subsection{Preconditioning and orthogonalisation}
\label{sec:orthogonalisation}

Our primary concern in the current implementation was to improve the convergence of the CT
equations. To achieve this, instead of using a diagonal
preconditioner as in (\ref{eq:diag1}), (\ref{eq:diag2}), we
updated the amplitudes through an exact Newton step. The simplest way
to define the Newton update is through the linear equation
\begin{align}
D^{p,\,v}_{s,\,y} \, \Delta A^{v}_{y} &= -R^{p}_{s} \label{eq:lineareq_a1} \\
D^{pq,\,vw}_{st,\,yz} \, \Delta A^{vw}_{yz} &= -R^{pq}_{st} \label{eq:lineareq_a2}
\end{align}
with
\begin{align}
D^{p,\,v}_{s,\,y} &=
\langle [[\bar{H}_{1,2}, a^{v}_{y}-a^{y}_{v}]_{1,2}, a^{p}_{s}-a^{s}_{p}]_{1,2} \rangle \label{eq:def_d1}\\
D^{pq,\,vw}_{st,\,yz} &=
\langle [[\bar{H}_{1,2}, a^{vw}_{yz}-a^{yz}_{vw}]_{1,2},
  a^{pq}_{st}-a^{st}_{pq}]_{1,2} \rangle \label{eq:def_d2}
\end{align}
We can interpret the  $D$ matrices  as the derivatives of the
residual or Hessians of the energy. However, eqns. (\ref{eq:def_d1}),
(\ref{eq:def_d2}) are
non-optimal as the search directions (i.e. the components of $A$) within the
first-order interacting space, namely
the  singly-external,
doubly-external, and semi-internal excitations
\begin{align}
(a^i_a - a^a_i) \Psi_0  \label{eq:singly_external} \\
(a^{ab}_{ij} - a^{ij}_{ab}) \Psi_0  \label{eq:doubly_external} \\
(a^{ak}_{ij} - a^{ij}_{ak}) \Psi_0  \label{eq:semi_internal}
\end{align}
generate a non-orthogonal and even linearly-dependent basis.
The large spread in eigenvalues of the
overlap of the first-order interacting basis (\ref{eq:singly_external}),
(\ref{eq:doubly_external}), (\ref{eq:semi_internal})
can then cause poor convergence of the linear equations (\ref{eq:lineareq_a1}) and
(\ref{eq:lineareq_a2}).

To remedy this,
we first orthogonalise the first-order interacting basis  by diagonalising the overlap matrix $S$  made of the
one, two, and three-particle density matrices,
\begin{align}
S_{i,\,j} &= \langle (a^{a}_{i} - a^{i}_{a})^\dagger (a^{a}_{j} - a^{j}_{a}) \rangle \nonumber \\
          &= \gamma^{i}_{j} \label{overlap_single_single} \\
S_{i,\,jkl} &= \langle (a^{a}_{i} - a^{i}_{a})^\dagger (a^{al}_{jk} - a^{jk}_{al}) \rangle \nonumber  \\
            &= \gamma^{il}_{jk} \label{overlap_single_semi} \\
S_{ijk,\,lmn} &= \langle (a^{ak}_{ij} - a^{ij}_{ak})^\dagger (a^{an}_{lm} - a^{lm}_{an})  \rangle \nonumber\\
              &= \delta_{kn} \gamma^{ij}_{lm} - \gamma^{ijn}_{lmk} \label{overlap_semi_semi} \\
S_{ij,\,kl} &= \langle (a^{ab}_{ij} - a^{ij}_{ab})^\dagger (a^{ab}_{kl} - a^{kl}_{ab})  \rangle \,\,\,\,\,(a\neq b)\nonumber\\
            &= \gamma^{ij}_{kl} \label{overlap_double_double}
\end{align}
and change to the orthogonalised excitation operators
$a^a_\mu$ and $a^{ab}_\nu$
\begin{align}
a^a_\mu &= S^{-1/2}_{\mu, i} (a^{a}_{i} - a^{i}_{a}) + S^{-1/2}_{\mu, ijk} (a^{ak}_{ij} - a^{ij}_{ak}) \label{orth_single} \\
a^{ab}_\nu &= S^{-1/2}_{\nu, ij} (a^{ab}_{ij} - a^{ij}_{ab})  \label{orth_double}
\end{align}



We can then solve the Newton equations (\ref{eq:lineareq_a1}), (\ref{eq:lineareq_a2})  in this orthogonalised representation. To do so, the
quantities $A$, $R$, and $D$ are
re-expressed in terms of $a^a_\mu$ and $a^{ab}_\nu$
\begin{align}
A &= \tilde{A}^a_\mu a^a_\mu + \tilde{A}^{ab}_\nu a^{ab}_\nu  \\
\tilde{R}^{a}_{\mu} &= \langle [\bar{H}_{1,2}, a^{a}_{\mu} - a^{\mu}_{a}]_{1,2} \rangle \\
\tilde{R}^{ab}_{\mu} &= \langle [\bar{H}_{1,2}, a^{ab}_{\mu} -
  a^{\mu}_{ab}]_{1,2} \rangle   \\
\tilde{D}^{a,\,b}_{\mu,\,\nu} &=
\langle [[\bar{H}_{1,2}, a^{b}_{\nu}-a^{\nu}_{b}]_{1,2}, a^{a}_{\mu}-a^{\mu}_{a}]_{1,2} \rangle \\
\tilde{D}^{ab,\,cd}_{\mu,\,\nu} &=
\langle [[\bar{H}_{1,2}, a^{cd}_{\nu}-a^{\nu}_{cd}]_{1,2},
  a^{ab}_{\mu}-a^{\mu}_{ab}]_{1,2} \rangle
\end{align}
The numbers of operators $a^a_\mu$ and $a^{ab}_\nu$ are $O(a^3e)$ and
$O(a^2e^2)$, respectively. Thus the additional cost of the transformation is
$O(a^6e)$ for the terms involving $a^a_\mu$ and $O(a^4e^2)$ for the
terms involving $a^{ab}_\nu$.  The diagonalisation of the overlap
matrix $S$ for the semi-internal and singly-external
(i.e.\ eqn.\ (\ref{overlap_single_single})-(\ref{overlap_semi_semi}))
requires a cost of $O(a^9)$. While the scaling of these steps is
relatively high, they are not expected to be a bottleneck
for systems where conventional CASSCF calculations can be performed
(for example, internally contracted CASPT2 also contains steps with
such cost \cite{AMRSW:1990:_caspt2, AMR:1992:_caspt2}). However, if we
were to use a large active space arising from e.g. a DMRG calculation,
a different algorithm should be used.

Let us consider now the condition number of $\tilde{D}$ and the
convergence characteristics of the Newton equations in the
orthogonalised representation.
If $D$ were formed \textit{without} any operator decomposition approximation, then
 $\tilde{D}$ would represent the true Hessian of the energy with respect to
 an orthogonal set of directions in the first-order interacting space. The condition number of $\tilde{D}$
 would then be governed by the excitation energy between the reference
 and excited states, which could be expected to be reasonable in most
 systems. Loosely speaking, we  can regard the improved condition
 number of $\tilde{D}$ as arising from the cancellation of small
 eigenvalues of $D$ by the
 large eigenvalues of $S^{-1/2}$. 
 However, such a cancellation is unstable,
if we  approximate   $D$ using  the operator
decomposition. 
Therefore to further improve the
 condition number of $\tilde{D}$  we
discarded those operators $a^a_\mu$ and $a^{ab}_\nu$
 which corresponded to small eigenvalues of
 $S$. The eigenvalue truncation thresholds are denoted hereafter as
$\tau_s$ for the singly external and semi-internal and
$\tau_d$ for the doubly external excitations. 
This limits the largest linear combination
 amplitude coefficients (e.g. $S^{-1/2}_{\mu, i}$) appearing in eqns. (\ref{orth_single}),
 (\ref{orth_double}) to $O(\tau_s^{-1/2})$ and $O(\tau_d^{-1/2})$
 respectively, preserving numerical stability in the amplitude equations.
Typically, we used  $\tau_s < 10^{-1}$  and  $\tau_d < 10^{-2}$. These cutoffs appear large because of the extreme
degeneracy of the first-order interacting space near equilibrium, and because of the incomplete removal of the poorly conditioned
components, due to the slight incompatibility (unstable cancellation) between the approximate Hessian and the overlap matrix in this space.
Linear
dependency is particularly strong near equilibrium because some of the active
orbitals which are being excited by $A$ have nearly zero occupancy.
However, the contribution of the neglected terms to the energy is small.
 For N$_2$ with the
cc-pVDZ basis, the size of the effective orthogonalised operator space was
410 ($R_{NN}=1.0$), 698 ($R_{NN}=1.6$), and 938 ($R_{NN}=3.0$)  indicating that over
50\% of the operator basis was truncated near equilibrium. In the dissociation region of the potential energy curves
studied here,  truncation 
 did not occur.


In our previous work, we encountered numerical difficulties in using singly
external excitation operators in conjunction with doubles,
i.e.\ for L-CTSD.  We now see that the reason is the linear dependency between singly external
and semi-internal excitations, which appears as non-zero overlap in
$S_{i,\,jkl}$ (eqn.\ (\ref{overlap_single_semi})). The
orthogonalisation fixes this issue, and thus we have used L-CTSD as the
standard L-CT model in this work. This should be naturally superior to L-CTD as it
includes  orbital relaxation and extra correlation such as
three- or higher-particle excitations from the direct product of
singles and doubles.

Using  the Newton update as described above,  we observed efficient convergence
in the CT amplitude equations. Typically only 10 Newton steps would be
required to converge the amplitudes in multireference calculations.
Convergence behavior of the amplitude equations in L-CTSD(CU) and L-CTSD(\MK) was
generally similar, but there were some cases where we could converge
the L-CTSD(\MK) but not  the L-CTSD(CU) calculations with the standard
truncation thresholds, for example at  $R_{NN} = 1 \AA$ for  N$_2$
(cc-pVTZ),  as discussed in Sec. \ref{sec:water_nitrogen}.

\subsection{Operator orthogonalisation with cumulant density matrix}
\label{sec:cumulant_orth}

Rather than using the exact three-particle density matrix for the
orthogonalisation procedure described above,  we could  also imagine using its  cumulant
decomposition in terms of the one- and two-particle density matrices, eqn.\ (\ref{eq:threecum}).


Table \ref{table:app3rdm} shows the
differences of the total energies computed using the amplitude
operators that are orthogonalised with the exact and approximate (cumulant)
three-particle density matrices for the N$_2$ potential energy curves
discussed in section \ref{sec:water_nitrogen}. With the truncation threshold $\tau_s = 10^{-1}$ and $\tau_d =
10^{-2}$ and using various basis sets, the energies from both
orthogonalisations were generally in good agreement within a deviation of 1.2
m\Ehartree.  However, at $R_{NN}=1.0\AA$ with the cc-pVTZ
basis set, the L-CTSD(MK) calculation with the cumulant based orthogonalisation
did not converge.



\begin{table}[t]
\begin{center}
\caption{The total energies (\Ehartree) of L-CTSD(\MK) for the bond
  breaking of the
  N$_2$ molecule with CAS$(6e, 6o)$ and various basis sets using the
  exact and approximate (cumulant) three-particle density matrices for
  orthogonalisation. }
\label{table:app3rdm}
\begin{tabular}{lrrr}
\hline
\hline
        & $1 \AA$  & $2 \AA$  & $3 \AA$  \\
\hline
\hline
6-31G \\
exact orthog.    & -109.04585 & -108.86156 & -108.84155 \\
cumulant orthog. & -109.04566 & -108.86220 & -108.84155 \\
diff (m\Ehartree)    &      +0.19 &      -0.64 &        0.0 \\
\hline
cc-pVDZ \\
exact orthog.    & -109.22774 & -108.98214 & -108.96009 \\
cumulant orthog. & -109.22752 & -108.98305 & -108.96009 \\
diff (m\Ehartree)    &      +0.22 &      -0.91 &        0.0 \\
\hline
cc-pVTZ \\
exact orthog.    & -109.33525 & -109.05709 & -109.03011 \\
cumulant orthog. &  not conv. & -109.05824 & -109.03011 \\
diff (m\Ehartree)    &            &      -1.15 &        0.0 \\
\hline
\hline
\end{tabular}
\end{center}
\end{table}

\section*{}
\bibliographystyle{aip}
\bibliography{ct-ref}

\end{document}